\documentclass[11pt]{article}
\usepackage{graphicx,cite,amsmath}

\renewcommand{\appendix}
        {
        \par
        \setcounter{section}{0}
        \setcounter{subsection}{0}
        \gdef\afterthesectionpunctdefault{:}
        \gdef\thesection{{Appendix \Alph{section}}}
        \renewcommand{\theequation}{\Alph{section}\arabic{equation}}
        \setcounter{equation}{0}
        }
\def\lsim{\hbox{\lower .8ex\hbox{$\, \buildrel < \over \sim\,$}}}
\def\gsim{\hbox{\lower .8ex\hbox{$\, \buildrel > \over \sim\,$}}}

\def\ds{\displaystyle}
\def\ver{\vec r \,}
\def\vrp{\vec{r\,}'}
\def\vs{\vec s\, }
\def\vr0{\vec{r}_0}
\def\t{\theta}
\def\tp{\theta '}
\def\ep{{\mbox{\large e}}}

\newfont{\ensmathquatorze}{msbm10 scaled 1400}
\newfont{\ensmathonze}{msbm10 scaled 1100}
\newfont{\ensmathdix}{msbm10}
\newfont{\ensmathneuf}{msbm10 scaled 833}
\newfont{\ensmathhuit}{msbm10 scaled 694}
\newfam\ensmathfam                        
\textfont\ensmathfam=\ensmathonze        
\scriptfont\ensmathfam=\ensmathdix       
\scriptscriptfont\ensmathfam=\ensmathhuit
\def\ensmf{\fam\ensmathfam\ensmathonze}         

\def\NN{{\ensmf N}}                 
\def\ZZ{{\ensmf Z}}                 

\begin{document}

\begin{center}

{\huge Diffractive corrections in the }\\

\

{\huge  trace formula for polygonal billiards }\\

\vspace{1.0cm}

{\Large E. Bogomolny, N. Pavloff and C. Schmit}
\end{center}

\vspace{0.5 cm}

\begin{center}
\noindent Laboratoire de Physique Th\'eorique
et Mod\`eles Statistiques\footnotemark,\\
\vspace{0.3 cm}
Universit\'e Paris Sud, b\^at. 100, F-91405 Orsay Cedex, France 

\vspace{2 cm}

{\bf Abstract}
\end{center}

	We derive contributions to the trace formula for the spectral density
accounting for the role of diffractive orbits in two-dimensional polygonal
billiards. In polygons, diffraction typically occurs at the boundary of a
family of trajectories. In this case the first diffractive correction to the
contribution of the family to the periodic orbit expansion is of order of the
one of an isolated orbit, and gives the first $\sqrt{\hbar}$ correction to the
leading semi-classical term. For treating these corrections Keller's
geometrical theory of diffraction is inadequate and we develop an alternative
approximation based on Kirchhoff's theory. Numerical checks show that our
procedure allows to reduce the typical semi-classical error by about two
orders of magnitude. The method permits to treat the related problem of
flux-line diffraction with the same degree of accuracy.

\vspace{3cm}
\begin{math}
\footnotetext[1]{Unit\'e Mixte de Recherche de l'Universit\'e Paris XI et du
CNRS (UMR 8626).}
\end{math}

\noindent PACS numbers:\hfill\break
\noindent 05.45.Mt ~ Semiclassical chaos (``quantum chaos'').\hfill\break
\noindent 03.65.Sq ~ Semiclassical theories and applications.\hfill\break
\noindent 42.25.Fx ~ Optics, diffraction and scattering.\hfill\break
\newpage

\section{Introduction}

	Two dimensional billiards play a central role in the domain of quantum chaos
because of the simpli\-ci\-ty of their classical dynamics and of the
relatively easy determination of their quantum spectrum. During the last 20
years they have been used as model systems for testing semiclassical trace
formulae (following Gutzwiller \cite{gut89} and Balian and Bloch \cite{bal72})
and random matrix theory (see e.g. \cite{boh91}).

	Amongst these systems, plane polygonal billiards have been subject of a long
lasting interest (see e.g. the review \cite{gut86})~: they have zero metric
\cite{sin76} and topological\cite{gal95} entropy, but their dynamical
properties range from integrable to possibly ergodic and mixing
\cite{cas99}, passing by the interesting group of pseudo-integrable systems
\cite{ric81}. Level correlation of integrable polygonal billiards display
inte\-res\-ting properties \cite{itz86}, not to speak of the case of
pseudo-integrable billiards whose level statistics are intriguingly related to
those of the Anderson model at the metal-insulator point \cite{kra97,bog99}.

	The present work is devoted to the detailed study of the trace formula in
polygonal billiards. Though the general method of deriving the trace formula
is well known \cite{gut89}, its application to polygonal plane billiards is
not straightforward. The main difficulty is the existence of important
corrections due to the diffraction on the corners of the billiard. This type
of correction was treated in Refs.~\cite{vat94,pav95,bru96} in the
framework of Keller's geometrical theory of diffraction \cite{kel62}. This
amounts to introduce in the trace formula new diffractive orbits that obey
the laws of classical mechanics everywhere except on singularities of the
potential where they are diffracted non-classically. The result of the
approach of Refs.~\cite{vat94,pav95,bru96} diverges when a diffractive
orbit is close to be allowed by classical mechanics~; this deficiency was
remedied in some special cases in Refs.~\cite{pri97,sie97}. Ref.~\cite{sie97}
studies corner diffraction in two dimensional billiards (not exclusively
polygons). It gives uniform formulae but is limited to single diffraction.
Ref.~\cite{pri97} treats diffraction by a circular disk inside a billiard. It
considers up to doubly diffractive orbits, but does not provide a uniform
approximation. In the present paper we extend these approaches and construct
improvements to the geometrical theory of diffraction in polygonal billiards.

	This type of corrections is made necessary in polygons because in these
systems the spatial extension of a family of orbits is often stopped by a
singularity of the frontier of the billiard~; as a result the generic
situation is that a diffractive orbit appears on the boundary of each family.
This trajectory is on the verge of being allowed by classical mechanics and
thus cannot be included in the trace formula in the framework of the
geometrical theory of diffraction. Hence in the following we devote a special
care to the treatment of diffractive periodic orbits lying on the boundary of
a family and of its repetitions. We give explicit formulae for the corrections
to the leading semi-classical term for the $n^{th}$ iterate of a family of
periodic orbits.

 We find in polygonal billiards a very rich variety of diffractive orbits.
Their contributions give $\sqrt{\hbar}$ corrections to the leading
semi-classical term in the trace formula and allow to compute the level
density with great precision. Numerical checks show that the typical
semiclassical error is reduced by one or two orders of magnitude.

	The paper is organized as follows. In Section \ref{section2} we briefly
present Keller's geometrical theory of diffraction and propose an alternative
approximation based on Kirchhoff's theory that is valid near the ``optical
boundary'' (the separation between allowed and forbidden classical
trajectories, in optics this occurs on the line separating light and shadow).
The simplicity of the method permits a straightforward generalization to the
case of diffraction by a flux line. The approximation established in Section
\ref{section2} is used to treat a large number of different types of
diffractive periodic orbits. We first consider corner diffraction. The
contribution of a diffractive orbit on the boundary of a periodic orbit family
is calculated in Section \ref{front}. This is a typical situation for
pseudo-integrable billiards. Special attention is given to the diffractive
partner of the $n$-fold repetition of a primitive periodic orbit. Section
\ref{deuxfront} is devoted to the study of diffractive orbits that are
simultaneously on the boundary of a family and on the frontier of the
billiard. Another type of diffractive orbits that belong to the boundaries of
two different families of periodic orbits is discussed in Section \ref{jump}.
Besides diffractive orbits lying exactly on an optical boundaries, there exist
orbits that are so close to an optical boundary that the geometrical theory
of diffraction cannot be applied. Such orbits are studied in Sections
\ref{updown} and \ref{section7}. All these special cases are necessary for a
careful description of the quantum density of states in pseudo-integrable
billiards. In Sec.~\ref{section8} we illustrate the flexibility of our
approach by presenting results for flux line diffraction in a rectangular
billiard. In this case, solving the question of diffraction on the optical
boundary amounts to treat the non-trivial problem of (multiple) forward
Aharonov-Bohm scattering. Finally we present our conclusions in Section
\ref{conclusion}. Some technical points are given in the Appendices. In
\ref{unif} a concise discussion of improvements to Keller's theory of
diffraction is given. In \ref{integrales} we discuss the computation of
certain trace integrals. In Appendices C and D we derive analytically explicit
expressions for important $n$-dimensional integrals.

\section{Diffractive Green function}\label{section2}

	In this Section we first present Keller's geometrical theory of diffraction
(putting the emphasis on corner diffraction) and then propose an alternative
approximation valid near the optical boundary (when Keller's approach fails)
for corner and flux-line diffraction.

\subsection{Geometrical theory of diffraction}

	One considers the different approximate contributions to the Green
function $G(\ver,\vrp,E)$ for two points $\vec{r}$ and $\vec{r\,}'$
in a polygonal billiard. The first is the semi-classical contribution which is
a sum over all possible classical trajectories going from $\vec{r\,}'$ to
$\vec{r}$. It is of the form :

\begin{equation}\label{e1}
G_{0}(\vec r , \vrp ,E)= 
\underset{\mbox{\tiny{classical}}}{\sum_{\vrp\to\vec r}}
\frac{\ds \exp\{ i (k L -\nu\pi/2-3\pi/4)\}}{\sqrt{\ds 8\pi k L}} \; ,
\end{equation}

\noindent where $L$ is the length of the classical path going from $\vrp$ to
$\vec r$ and $\nu$ is twice the number of specular reflections along that path
(we consider Dirichlet boundary conditions). We use units such that the energy
is related to the wave-vector through $E=k^2$.

	There are other contributions to $G$ that correspond to diffractive orbits
experiencing specular reflections on the frontier of the billiard and also
non-classical bounces on the diffractive corner. In the framework of Keller's
geometrical theory of diffraction (see e.g. \cite{kel62}) a such orbit with a
single diffractive bounce contributes to the Green function with a term

\begin{equation}\label{e2}
G_{1d}(\vec r,\vrp,E)=G_{0}(\vr0,\vrp,E)\,
{\cal D}(\t,\tp)\, G_{0}(\vec r,\vr0,E) \; ,
\end{equation}

\noindent where $\vr0$ is the position of the diffractive apex and ${\cal D}$
is a diffraction coefficient depending on the interior angle $\gamma$ of the
polygon at $\vr0$ and on the incoming (outgoing) angle $\tp$ ($\t$) of the
diffractive trajectory with the boundary. The explicit expression of ${\cal
D}$ for corner diffraction reads \cite{kel62}~:

\begin{equation}\label{e3}
{\cal D}(\t,\tp) = \sum_{\sigma , \eta = \pm 1} \, {\cal D}_{\sigma ,
\eta}(\t,\tp) 
\qquad \mbox{where} \qquad
{\cal D}_{\sigma , \eta} (\t,\tp) = \frac{\ds 1}{\ds N}
\frac{\ds \sigma\, \eta}
{\ds \tan\left(\frac{\phi_\sigma+\eta\pi}{\ds 2N}\right)}
\; , \end{equation}

\noindent where $N=\gamma/\pi$ and $\phi_\sigma=\tp-\sigma\t$, (with $\t$ and
$\tp$ in $[0,\gamma]$). Although formula (\ref{e2}) is attached to the name of
Keller, the idea of treating diffraction as arising from a kind of reflection
on the edge has a long history which goes back to Young (see Chap.~44 of
\cite{som54} and Chap.~8.9 of \cite{bor87}). We also note here that the
repercussion of diffractive periodic orbits on the spectrum seems to have been
first worked out within the geometrical theory of diffraction by Durso in 1988
\cite{dur88}.

	Formula (\ref{e2}) can be generalized to treat multiple diffraction. One has
then several diffraction coefficients ${\cal D}_1, {\cal D}_2, \ldots $, one
for each diffractive bounce, and between each diffractive bounce a
semi-classical propagation described by a Green function of type (\ref{e1}).
When diffractive trajectories are taken into account in the trace formula, one
is lead to consider diffractive periodic orbits which contributions to the
level density $\rho(E)$ are of the form \cite{vat94,pav95,bru96}~:

\begin{equation}\label{e3bis}
\rho(E) \leftarrow
{\ds L_d\over\ds 2 \pi k} 
\left\{ \prod_{j=1}^{n} {\ds {\cal D}_j\over\ds\sqrt{\ds
8\pi k \ell_j}} \right\} \, \cos (k L_d -\nu_d\pi/2-3 n \pi/4) \; .
\end{equation}

	In (\ref{e3bis}) and in many instances below, when writing explicitly the
contribution of a periodic orbit (classical or diffractive) to the level
density, we put an arrow in direction of $\rho(E)$ for indicating that this is
one contribution amongst many others. In the above expression $\ell_1 , ...,
\ell_n$ are the lengths along the orbit between two diffractive reflections.
$\ell_1+...+\ell_n=L_d$ is the total length of the diffractive periodic orbit.
$\nu_d$ is the Maslov index of the diffractive orbit, i.e. twice the number of
specular reflections. Repetitions of a primitive diffractive orbit appear as a
special case of (\ref{e3bis}) ; in this case however, in the first factor
$L_d/(2 \pi k)$ of the r.h.s. of (\ref{e3bis}), $L_d$ should be understood as
the primitive length of the orbit.

\

	We recall that in a polygon, the contribution of an isolated periodic orbit
to $\rho(E)$ is of the form $-\ell/(4\pi k) \cos(k\ell)$ (for a primitive
orbit of length $\ell$). Thus Eq.~(\ref{e3bis}) shows that the contribution of
a typical diffractive periodic orbit with $n$ diffractive bounces is of order
${\cal O}(k^{-n/2}$) compared to that of an isolated periodic orbit. We will
study below special configurations where this is not the case and where
diffractive orbits have the same weight as isolated periodic ones. We first
need to discuss the range of validity of the geometrical theory of diffraction
and to define approximations alternative to (\ref{e2}).

\subsection{In vicinity of an optical boundary}\label{vob}

\begin{figure}[thb]
\begin{center}
\includegraphics*[width=7cm,bbllx=45pt, bblly=33pt, bburx=400pt,
bbury=380pt]{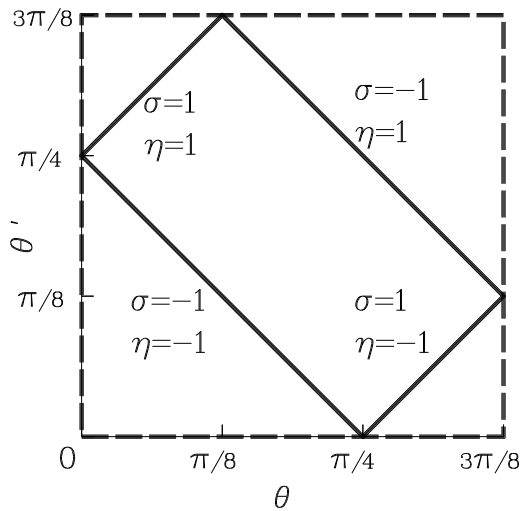}
\end{center}
\caption{\small Solid lines~: location of the angles $\t$ and $\t'$ for which
expression (\ref{e3}) of ${\cal D}$ diverges when $\gamma=3\pi/8$.
Near each line are indicated the values $\sigma$ and $\eta$ of
the divergent ${\cal D}_{\sigma,\eta}$.
Dashed lines~: location of the angles $\t$ and $\t'$ for which
expression (\ref{e3}) of ${\cal D}$ is zero. These correspond to $\t$ (or
$\t'$) $=0$ or $\gamma$.}
\label{opt_bound}
\end{figure}

	The approximation defined by Eqs. (\ref{e2},\ref{e3}) fails when the
diffractive bounce at $\vr0$ is ``almost allowed'' by classical mechanics. In
that limit the trajectory lies on what is called an ``optical boundary'' in
the literature, and the coefficient ${\cal D}$ diverges. This failure of
Keller's approximation can be intuitively understood by noting that
Eq.~(\ref{e2}) gives a contribution to the Green function of order ${\cal
O}(k^{-1})$ whereas in the limit that the diffractive orbit becomes allowed by
classical mechanics it has to contribute at order ${\cal O}(k^{-1/2})$ as any
classical trajectory. Hence in this limit Eq. (\ref{e2}) cannot hold. We study
below a triangle with a diffractive corner of opening angle $\gamma=3\pi/8$.
In this case, one can easily check geometrically that the diffractive orbit
coincides with a classical trajectory if the angles $\t$ and $\tp$ lie on one
of the lines of $[0,\gamma]^2$ shown on Fig.~\ref{opt_bound}. This can be also
checked algebraically from formula (\ref{e3})~: each of the four lines of Fig.
\ref{opt_bound} correspond to divergence of one of the coefficients ${\cal
D}_{\sigma , \eta}$.

\begin{figure}[thb]
\begin{center}
\includegraphics*[width=12.0cm,bbllx=35pt, bblly=137pt, bburx=575pt,
bbury=465pt]{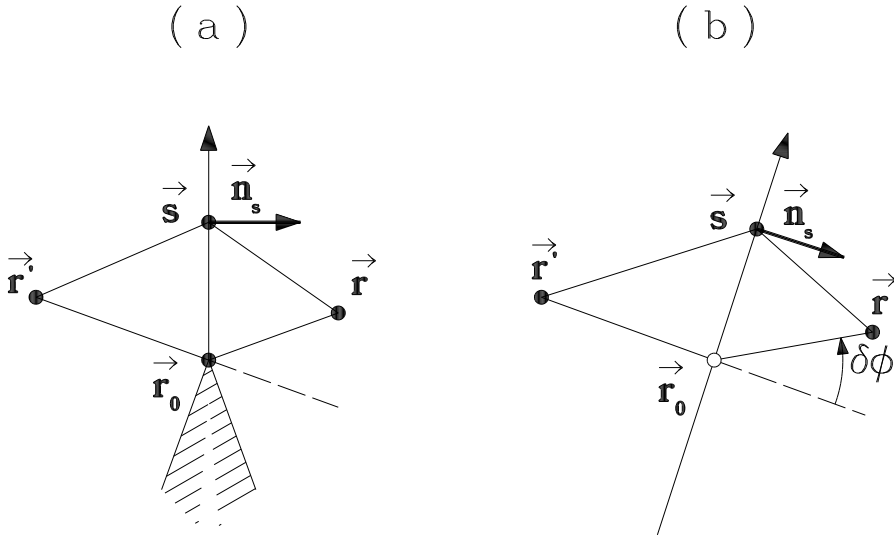}
\end{center}
\caption{\small Graphical representation of the notations of Eqs.~(\ref{e4})
and (\ref{e5}). Part (a) refers to corner diffraction and part (b) to flux
line diffraction. $\vr0$ is the diffractive point. The dashed line issued from
$\vr0$ is the optical boundary on which the geometrical theory of diffraction
fails. In part (a) the integration along $s$ is stopped on the apex at $\vr0$.
This is not the case in part (b). There however, near the optical boundary
$|\delta\phi| \ll \pi$ and the integrand of (\ref{e5}) contributes with a
phase that is approximatively $-\alpha\pi$ if $s>0$ and $\alpha\pi$ if
$s<0$.}
\label{fig_gene}
\end{figure}

	For corner diffraction, after the work of Pauli \cite{pau38}, several
uniform approximations have been derived which correct the drawbacks of
Eq.~(\ref{e2}). We recall one of these in \ref{cd}. In this paper we use a
simple approximation to the exact formula valid only near the optical
boundary. Let's consider that the trajectory lies near the optical boundary
defined by one of the four couples $(\sigma,\eta)$~; then our approximation for
the total Green function (semi-classical plus diffractive) reads~:

\begin{eqnarray}\label{e4}
G_1(\vec r,\vrp,E)& =&  -2 \int_0^{+\infty}\!\!\!ds\, G_{0}(\vs,\vrp,E) 
\; \vec{n}_s \cdot \vec{\nabla}_{\vs} G_{0}(\vec r,\vs,E) \nonumber\\
& & \nonumber\\
& +&G_{0}(\vr0,\vrp,E)\,
{\cal D}_{reg} (\t,\tp)\, G_{0}(\vec r,\vr0,E) \; ,
\end{eqnarray}

\noindent where the locus of points $\vs$ is an arbitrary half-line separating
$\vrp$ and $\vec r$ and issued from $\vr0$ (at $s=0$); $\vec{n}_s$ is a vector
normal to the $s$ axis and oriented towards $\vec{r}$, see
Fig.~\ref{fig_gene}a. ${\cal D}_{reg}$ is the non-divergent part of the
diffraction coefficient (i.e. the sum of all the ${\cal D}_{\sigma , \eta}$'s
but the divergent one). The diffractive Green function (analogous to
(\ref{e2})) is defined from Eq.~(\ref{e4}) by the difference~:
$G_{1d}=G_1-G_0$.

	Eq. (\ref{e4}) is a simple Kirchhoff approximation to the Green function
(with Keller-type corrections) which is to be used within the semiclassical
approximation (this is illustrated at length in the following sections). We
show in \ref{unif} how it can be derived starting from a more elaborate
approach. Eq.~(\ref{e4}) is exact in the limit that the classical path from
$\vrp$ to $\vec r$ lies on an optical boundary. It is designed to remedy
the divergence of the geometrical theory of diffraction and it is not a
uniform approximation to the Green function~: far from the optical boundary
characterized by $\sigma$ and $\eta$, it yields a result basically of the form
(\ref{e2}) without however the correct form of the coefficient ${\cal
D}_{\sigma,\eta}$ (this is in accordance with the known aspects of Kirchhoff
approximation, see e.g. \cite{som54}). This should not be considered as a
limitation of the approach~: we show below (Secs. \ref{updown} and
\ref{section7}) that it is a simple matter to ``uniformize'' the result
derived from (\ref{e4}) when necessary. Compared to the uniform expression
(\ref{a1}), Eq.~(\ref{e4}) has the important advantage of being easily
extended to treat multiple diffraction near the optical boundary
(Eq.~(\ref{f2})). In the following sections approximation (\ref{e4}) will
allow us to incorporate non-standard diffractive contributions in the trace
formula.

\

	Formula (\ref{e4}) can be extended to treat the case of diffraction by a
flux line. This problem bears important similarities with corner diffraction.
In some respect it can be considered as simpler, because for an initial point
$\vrp$, the diffractive point (the Aharonov-Bohm flux line) is associated with
a single diffractive boundary~: the forward direction. This is the reason why
formula (\ref{e5}) below -- which is the analogous of Eq.~(\ref{e4}) --
comprises only a Kirchhoff contribution and no Keller-like correction.

	We consider a particle of charge $q$ and a flux line located on point $\vr0$
such that the magnetic field is $\vec{B}=\Phi \delta(\vec{r}-\vr0\,) \hat{z}$.
The only relevant parameter is the ratio $\alpha$ of the flux $\Phi$ with the
quantum of flux $\alpha = q\Phi /(2\pi\hbar c)$ and one can restrict oneself
to $0<\alpha<1$. The Kirchhoff approximation for the total Green function is
(see the derivation in \ref{fld})~:

\begin{equation}\label{e5}
G_1(\vec r,\vrp,E) =  
-2 \int_{-\infty}^{+\infty}\!\!\!ds\, G_{0}(\vs,\vrp,E) 
\; \vec{n}_s \cdot \vec{\nabla}_{\vs} G_{0}(\vec r,\vs,E)
\, \ep^{\ds i\alpha \Delta\phi(s)}\; ,
\end{equation}

\noindent where the locus of points $\vec{s}$ is an
arbitrary line separating $\vrp$ and $\vec{r}$ and going through $\vr0$ (at
$s=0$), and $\Delta\phi(s)$ is the angle covered by the path going from $\vrp$
to $\vec{s}$ and then to $\vec{r}$. If $\delta\phi$ is the angle between
$\vec{r}-\vr0$ and $\vr0-\vrp$ (i.e. the departure from the optical boundary)
then $\Delta\phi(s)=\delta\phi-\pi\,\mbox{sgn}\,(s)$. Of course, in this
procedure, the orientation of the axis $(\vr0,s)$ is not arbitrary. Our choice
of $\Delta\phi(s)$ corresponds to an orientation such as presented on
Fig.~\ref{fig_gene}b.

	Eq. (\ref{e5}) has a simple physical interpretation~: the phase accumulated
by a trajectory depends on the sense of rotation of the circuit around the
flux line. As Eq.~(\ref{e4}) it is only valid near the optical boundary, but
it is easily generalized to multiple diffraction in the forward direction~:
i.e. it allows to treat the problem of multiple forward Aharonov-Bohm
diffusion. We illustrate this property in Sec.~\ref{section8}.

\section{A diffractive orbit on the frontier of a family}\label{front}

\begin{figure}[thb]
\begin{center}
\includegraphics*[width=8cm,bbllx=60pt, bblly=170pt, bburx=370pt,
bbury=320pt]{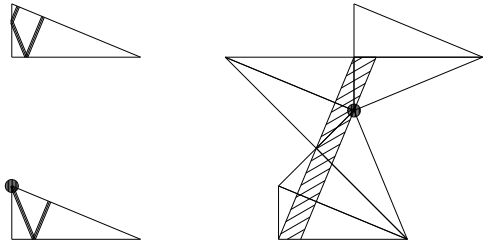}
\end{center}
\caption{\small Representation of a periodic orbit which is part of a family
in the triangle $(\pi/2,\pi/8,3\pi/8)$ (upper left triangle)
and of the diffractive orbit which is on the
boundary of the family (lower left triangle). The plot to the right
represents the family after 
unfolding. The area occupied by the family is shaded and the
diffractive point on its boundary is marked with a black dot (as in the
lower left triangle). The diffractive orbit in the lower left triangle
appears as the right boundary of the unfolded trajectory.}
\label{frontiere1}
\end{figure}

	A typical occurrence of diffractive orbits is at the boundary of a
family of trajectories. The width of a beam of classical orbits is limited
by a singularity of the frontier of the billiard. Such a case is illustrated
by the example shown in Fig.~\ref{frontiere1}. Note that this is not the
only possible type of boundary of a family. It may happen that the family
stops on a non-diffractive corner (a corner with opening angle of the
type $\pi/n$ with $n\in\NN$). This is the case for one of the boundaries of
the family displayed in Fig.~\ref{frontiere1}. The frontier of the family
may also be one of the frontiers of the billiard, this is illustrated on
Fig.~\ref{frontiere3}. We will also study below (Section \ref{deuxfront}) as
mixed case where the boundary of the family only partly coincides with the
frontier of the billiard.

	On Figure \ref{frontiere1} we have represented the family by one of
its member (upper left triangle). The boundary of the family is shown in the
lower left triangle. Also, to the right of the plot,
instead of representing the orbit by a series of segments
bouncing off the frontier of the billiard, we have represented
it by a unique
straight segment where the reflection on each edge is replaced by
continuing the path into a reflection of the enclosure.
This procedure is called ``unfolding the trajectory''.

	The diffractive orbit on the boundary of a family appears as a correction to
the contribution of the family and of its repetitions. Its contribution to the
trace formula cannot be evaluated from the geometrical theory of diffraction,
because its coefficient ${\cal D}$ is infinite. However, since the diffractive
orbit is exactly on an optical boundary, it can be described by using
Eq.~(\ref{e4}). The contribution of an orbit to the level density is evaluated
in the framework of a semiclassical periodic expansion (see e.g. Refs.
\cite{bal72,gut89}) ~: the level density being related to the Green function
through the trace $\rho(E)=-(1/\pi)\,\mbox{Im}\, \int d^2r\,
G(\ver,\ver,E+i0^+)$, the Green function is approximated in vicinity of a
periodic orbit (here it will be done by using Eq. (\ref{e4}) but other
approximations will be used below) and the trace integral is evaluated within
a saddle phase approximation near the saddle corresponding to the periodic
orbit considered.

\

	$\bullet$ The diffractive correction to the first iterate of a family is
very simple~: the trace of the first diffractive contribution to the Kirchhoff
Green function in (\ref{e4}) is zero and only the term with a regular
Keller-type diffraction appears. Hence, if the family has a length $\ell$ and
occupies on the billiard an area ${\cal A}$, then its total contribution
(semi-classical plus diffractive) to $\rho(E)$ is simply~:

\begin{equation}\label{f1}
\rho(E) \leftarrow
{\ds {\cal A}\over\ds 2 \pi}
\frac{1}{\sqrt{\ds 2\pi k \ell}}
\, \cos (k \ell -\pi/4)+
\frac{\ell}{2\pi k}
\frac{\ds {\cal D}_{reg}}{\sqrt{\ds 8\pi k \ell}}
\, \cos (k \ell -\nu_d\pi/2 - 3 \pi/4) \; .
\end{equation}

	The first part of the r.h.s. of (\ref{f1}) is the usual contribution
of a family of periodic orbits in two dimensions. The second part is of the
form (\ref{e3bis})~: it comes from the regular Keller contribution in
(\ref{e4}).

\

	In order to test the validity of our approach, we have compared our
analytical results with the spectrum determined numerically in a triangular
billiard with angles $(\pi/2,\pi/8,3\pi/8)$. Note that this is not a generic
polygonal billiard~: its classical mechanics is pseudo-integrable and
furthermore, it belongs to the set of ``Veech billiards'' \cite{veech}. We
choose these systems because they simplify the geometrical computations~:
Veech billiards have the amusing property that there exists only a finite
number of possible areas occupied by a family of periodic orbits. In the
triangle we study, one can show that there are only three possible areas~:
${\cal A}=1/\sqrt{2}$, $(\sqrt{2}+1)/2$ or $(\sqrt{2}-1)/2$ (we take the
hypotenuse of the triangle as unit length). We emphasize however that the
formulae obtained in the present paper are quite general.

 The numerical spectrum was obtained by expanding the wave function near the
angle $\pi/8$ in ``partial waves'' which are Bessel functions times a
sinusoidal function of the angle~: $\psi(r,\theta) = \sum_{m=1}^{m_{max}}
J_{8m}(kr) \sin(8 m \theta)$. This automatically fulfills the Dirichlet
conditions on the two faces of the billiard that meet at the corner with
opening angle $\pi/8$. The boundary condition on the remaining face is
enforced in a manner identical to the improved point matching method presented
in \cite{sch91}. This results in a secular equation whose solutions are the
eigenlevels of the system. We have tested the numerical stability of our
procedure by varying the number $m_{max}$ of partial waves. We have computed
the first 20 000 eigenlevels and we have checked that they were determined
with an accuracy of the order of 1/1000 of the mean level spacing.

	The agreement with the numerically determined spectrum can be
checked by studying the regularized Fourier transform on the level density~:

\begin{equation}\label{f1bis}
F(x)= \sqrt{\frac{\ds\beta}{\ds \pi(\Delta k)^2}}\,
\int_{k_{min}}^{k_{max}}\!\! dE\, \rho(E) \,\ep^{i k x}\,
\ep^{-\beta \left(\frac{k-k_{av}}{\Delta k}\right)^2}
\; .\end{equation}

	In (\ref{f1bis}) $k_{min}$ and $k_{max}$ are the lower and upper boundary
of a window of the spectrum (typically $k_{min}$ is the first eigen-level,
and $k_{max}$ the 5000$^{th}$ one);
$k_{av}=(k_{max}+k_{min})/2$ and $\Delta k=(k_{max}-k_{min})/2$. $\beta$
is a dimensionless regularizing parameter (typically $\beta=5$). If
$k_{min}=0$, $k_{max}\to +\infty$ and $\beta=0$, $F(x)$ is a series of
delta peaks centered on the lengths of the classical and diffractive
periodic orbits. 

\begin{figure}[thb]
\begin{center}
\includegraphics*[height=6cm,bbllx=28pt, bblly=90pt, bburx=525pt,
bbury=555pt]{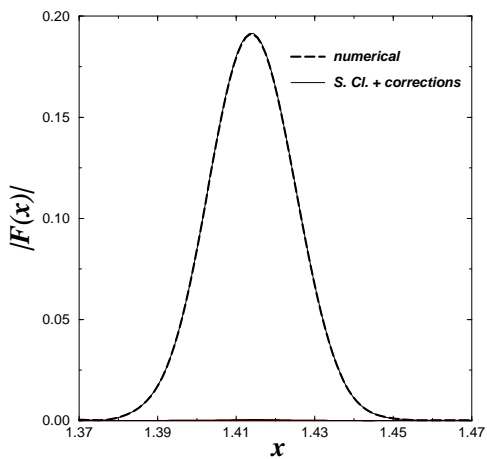}
\end{center}
\caption{\small Comparison of the numerical $|F(x)|$ (dashed line) with the
result from Eq. (\ref{f1}) (solid line). The two curves cannot be
distinguished, this will also occur in all the following plots of the same
type. We have taken here $\beta=5$, $k_{min}$ and $k_{max}$ being respectively
the first and the $5000^{th}$ level. The peak corresponds to the length of the
family shown in Fig.~\ref{frontiere1}, $\ell=\sqrt{2}$ (in all the text, the
hypotenuse of the triangle is chosen as unit length). The modulus of the
difference between the numerical and analytical values of $F(x)$
is also plotted in this figure, but is barely seen on this scale
(its largest value is $5\times 10^{-4}$). The usual semi-classical
contribution (with only the first term of the r.h.s. of (\ref{f1})) gives
instead an error of about $10^{-2}$.}
\label{fig1}
\end{figure}

	The comparison of the result of Eq.~(\ref{f1}) with the numerical datas is
shown in Fig.~\ref{fig1} for the family of Fig.~\ref{frontiere1}. The
agreement is excellent. Note that in this figure (and in the followings of the
same type) we compare different estimates for $|F(x)|$, but we also plot the
{\it modulus of the difference} between the numerical $F(x)$ and our
analytical formula, which is a strong test of accuracy. Note also that
for avoiding spurious sources of discrepancies with the numerical
result we compute the integral (\ref{f1bis}) numerically even
when we use an analytical expression for $\rho(E)$ (this corresponds to what
we still call the analytical $F(x)$).

\

	$\bullet$ We now concentrate on the second iterate of the family. Its
contribution is less trivial than (\ref{f1}) and more generic~; hence we will
present the computation in some details. One has here to consider double
diffraction near the optical boundary. Eq.~(\ref{e4}) is generalized to double
(and in a similar fashion to multiple) diffraction~:

\begin{eqnarray}\label{f2}
G_2(\vec r,\vrp,E)& =&  -2 \int_0^{+\infty}\!\!\!ds\, G_{1}(\vs,\vrp,E)\, 
\vec{n}_s \cdot \vec{\nabla}_{\vs} G_{0}(\vec r,\vs,E) \nonumber\\
& & \nonumber\\
& +&G_{1}(\vr0,\vrp,E)\,
{\cal D}_{reg} (\t,\tp)\, G_{0}(\vec r,\vr0,E) \; ,
\end{eqnarray}

\noindent where $G_2$ is the total (semi-classical plus diffractive) Green
function.

\begin{figure}[thb]
\begin{center}
\includegraphics*[width=8cm,bbllx=40pt, bblly=140pt, bburx=565pt,
bbury=450pt]{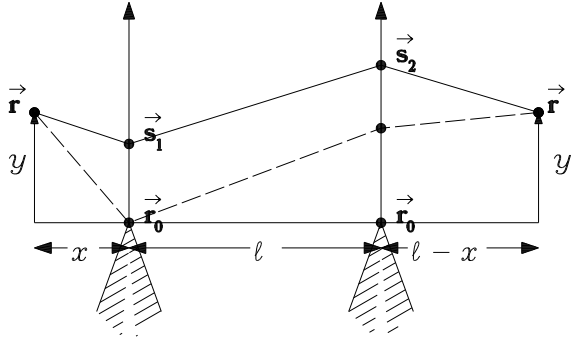}
\end{center}
\caption{\small Schematic representation of the different contributions to the
Green function $G_2$ near a diffractive periodic orbit on the boundary of the
second iterate of a family. In this plot (and in the followings of the same
type) the trajectory is represented after unfolding and the shaded areas are
zones forbidden by classical mechanics. The path represented by a solid line
going from $\vec r$ to $\vec s_1$, $\vec s_2$ and $\vec r$ contributes to the
leading Kirchhoff term in the first integral of the r.h.s. of (\ref{f2}). The
dashed path contributes at next order (it has one ``Keller bounce'' at $\vr0$
with diffraction coefficient ${\cal D}_{reg}$). Its contribution to the Green
function corresponds to the second term in (\ref{f5}).}
\label{frontiere2}
\end{figure}

	When unfolding the trajectory (as done for instance in
Fig.~\ref{frontiere1}) near the diffractive boundary of the family, one is
lead to consider contributions such as presented in Fig.~\ref{frontiere2}. On
that figure the position of a point $\vec r$ in vicinity of the diffractive
trajectory on the boundary of the family is defined by coordinates $x$ and
$y$. $x$ is a coordinate along the orbit ($0\le x\le \ell$) and $y$ a is
transverse coordinate.

	The leading order contribution of $G_2$ to the level density
is the usual contribution of the second
repetition of a family of periodic orbit. It is obtained by simply making
the approximation $G_2\approx G_{0}$ and it is of the form~:

\begin{equation}\label{f2bis}
\rho(E )\leftarrow 
{\ds {\cal A}\over\ds 2 \pi}
\frac{1}{\sqrt{\ds 2\pi k L}}
\, \cos (k L -\pi/4) \; .\end{equation}

	Here and in the following of this section $L=n\,\ell$ is the total length of
the trajectory, $\ell$ is the primitive length, and $n$ is the repetition
number (here $n=2$).

	The diffractive corrections
to (\ref{f2bis}) are included into $\rho(E)$ through the following 
trace~:

\begin{equation}\label{f2ter}
\rho(E)\leftarrow
-\frac{1}{\pi}\,\mbox{Im}\,\int d^2 r\,  
\left( G_2(\ver, \ver, E) -  G_{0}(\ver, \ver, E) \right)
\; .
\end{equation}

	From the expression (\ref{f2}) of $G_2$ one obtains the first order
contribution to $G_2(\ver,\ver,E)-G_{0}(\ver,\ver,E)$ under the form~:

\begin{eqnarray}\label{f3}
\frac{\sqrt{\ds k}\, \ep^{\ds i k L+3i\pi/4}}
{2\, (2\pi)^{3/2}\sqrt{\ds x \ell (\ell-x)}}\!\!\!\! &\!\!\!\!  & \left[
\int_0^{+\infty}\!\!\!\!\!\!ds_1
\int_0^{+\infty}\!\!\!\!\!\!ds_2\,
\ds\ep^{\ds i\frac{\ds k}{\ds 2}
\left\{\ds \frac{\ds (s_1-y)^2}{\ds x}+\frac{\ds(s_2-s_1)^2}{\ds \ell}
+\frac{\ds (y-s_2)^2}{\ds \ell-x}\right\}} \right. \nonumber\\
& & \nonumber\\
 & - \!\!\!\!& \Theta(y) 
\left. \int_{-\infty}^{+\infty}\!\!\!\!\!\!ds_1
\int_{-\infty}^{+\infty}\!\!\!\!\!\!ds_2\,
\ds\ep^{\ds i\frac{\ds k}{\ds 2}
\left\{\ds \frac{\ds (s_1-y)^2}{\ds x}+\frac{\ds(s_2-s_1)^2}{\ds \ell}
+\frac{\ds (y-s_2)^2}{\ds \ell-x}\right\}}
\right]
\; .\end{eqnarray}

		This contribution corresponds to the main Kirchhoff term in (\ref{f2})
(i.e. to the solid path from $\vec r$ to $\vec r$ in Fig.~(\ref{frontiere2}))
from which the semi-classical contribution has been removed when it exists,
i.e. when $y>0$ (this semi-classical contribution has been written as the
double sum $\int_{-\infty}^{+\infty}ds_1 \int_{-\infty}^{+\infty}ds_2$ in
(\ref{f3})). In the expression (\ref{f3}) one has made the hypothesis
$|y-s_1|,|s_1-s_2| ,|y-s_2| \ll x,\ell$. Thus, for instance,
$|\vec{s}_2-\vec{s}_1|\simeq \ell+(s_2-s_1)^2/(2\ell)$. The above expression
has to be inserted into Eq. (\ref{f2ter}), i.e. integrated transverse to the
orbit (along $y$) and longitudinally (along $x$). This is done in
\ref{integrales} and the resulting contribution to the level density is~:

\begin{equation}\label{f4}
\rho(E)\leftarrow -\frac{\ell}{8\pi^2 k} \, \cos (k L)
\; .\end{equation}

	This shows that the main diffractive correction to the contribution
of the second iterate of a family is of order of the one of an isolated
periodic orbit. Such non-generic contributions in vicinity of a family
have already been studied in a slightly different context
in Ref.~\cite{pri97}.

\

	For a better agreement with numerical data, one needs to include also
the next order correction to (\ref{f4}) in the level density. This is done
by including mixed Kirchhoff-Keller contributions in the Green function
(\ref{f2}), such as described by the path
represented in Fig.~(\ref{frontiere2}) by a dashed line. Along that path, the
first diffraction at $\vr0$ is of Keller-type (involving a coefficient
${\cal D}_{reg}$), and the second one of Kirchhoff type (with an integral
along $s_2$). One has two contributions,
one for each possible location of Keller diffraction (a
single one being shown in Fig.~(\ref{frontiere2})). The relevant
contribution to $G_2$ are now of the form~:

\begin{equation}\label{f5}
\frac{{\cal D}_{reg}\, \ep^{\ds i k L+i\pi/4-i\nu_d\pi/2}}
{4\, {\sqrt{\ds k}\,(2\pi)^{3/2}\sqrt{\ds x \ell (\ell-x)}}}
\left[
\int_0^{+\infty}\!\!\!\!\!\!ds_1\,
\ep^{i\frac{k}{2}
\left\{ \frac{(s_1-y)^2}{x}+\frac{s_1^2}{\ell}
+\frac{y^2}{\ell-x}\right\}}
+
\int_0^{+\infty}\!\!\!\!\!\!ds_2\,
\ep^{i\frac{k}{2}
\left\{\frac{y^2}{x}+\frac{s_2^2}{\ell}
+\frac{(y-s_2)^2}{\ell-x}\right\}}
\right]
\; .\end{equation}

	The integral of this expression is computed in \ref{integrales} (Eq.
(\ref{b10})). It
yields the next correction to (\ref{f4}) which is of the form~:

\begin{equation}\label{f6}
\rho(E )\leftarrow 
\frac{\ell}{2 \pi k}\, \frac{{\cal D}_{reg}}{\sqrt{8\pi k L}}
\, \cos (k L -\nu_d\pi/2 - 3\pi/4)
\; .\end{equation}

	In (\ref{f6}) $\nu_{d}$ is the Maslov index of the orbit corresponding to
the dashed path in Fig.~\ref{frontiere2}. If $\sigma$ is the relevant index
near the optical boundary considered (i.e. the one for which ${\cal
D}_{\sigma,\eta}$ in Eq.~(\ref{e3}) diverges), one can show that $\exp
\{-i\nu_{d}\pi/2\}=\sigma$.

\

\begin{figure}[thb]
\begin{center}
\includegraphics*[height=6cm,bbllx=28pt, bblly=90pt, bburx=512pt,
bbury=555pt]{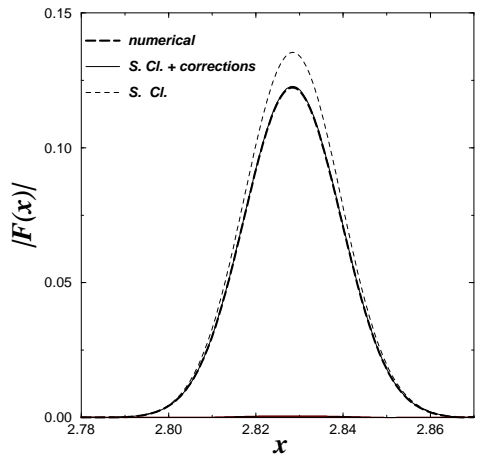}
\end{center}
\caption{\small Same as Fig.~\ref{fig1} for the second iterate of the family.
The solid curve (labelled ``S. Cl. + corrections'') corresponds to Eqs.
(\ref{f2bis},\ref{f4},\ref{f6}). In this plot and in the followings of the
same type, we represent with a thin dashed line (denoted ``S. Cl.'') the usual
semiclassical result without diffractive corrections (which corresponds here
to Eq.~(\ref{f2bis}) alone). The modulus of the difference between the
numerical and analytical $F(x)$ is also plotted in this figure, but is barely
seen on this scale~; its largest value is $5\times 10^{-4}$, whereas the usual
semi-classical approach gives an error of $1.4\times 10^{-2}$. }\label{fig2}
\end{figure}

\begin{figure}[thb]
\begin{center}
\includegraphics*[width=8cm,bbllx=60pt, bblly=230pt, bburx=345pt,
bbury=330pt]{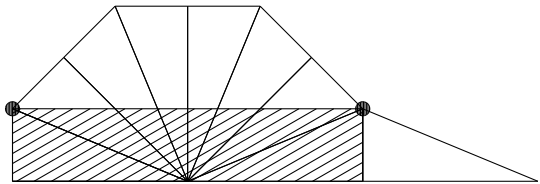}
\end{center}
\caption{\small Representation of a family of periodic orbits in the triangle
$(\pi/2,\pi/8,3\pi/8)$ by the technique of unfolding. The family has a
length $\ell=2\cos\pi/8$. Its area is shaded and the
diffractive point on its boundary is marked with black dots.}
\label{frontiere3}
\end{figure}

	There is a last correction to (\ref{f6}), purely of Keller-type, giving a
contribution of order ${\cal O}(k^{-2})$, but the contributions (\ref{f2bis}),
(\ref{f4}) and (\ref{f6}) already give a very good description of the Fourier
transform of the spectrum. This can be checked in Fig.~\ref{fig2} for the
second iterate of the family drawn in Fig.~\ref{frontiere1}.

\

	A simple remark is in order here~: Eq.~(\ref{f2bis}) is actually the
first term of an expansion in $k^{-1}$ (or equivalently in $\hbar$). The
magnitude of the next correction can be estimated by the following argument~:
the exact Green function in an infinite wedge can be expressed in
terms of a Hankel function (with diffractive corrections unimportant for the
present discussion) and this yields instead of Eq.~(\ref{f2bis}) to something
like

\begin{eqnarray}\label{f6bis}
\rho(E)& \leftarrow & \frac{{\cal A}}{4\pi} J_0(k L)\nonumber\\
       & \approx & {\ds {\cal A}\over\ds 2 \pi}
\frac{1}{\sqrt{\ds 2\pi k L}}
\, \cos (k L -\pi/4) + 
\frac{\ell}{2 \pi k}\, \frac{{\cal A}/(8\,L^2)}{\sqrt{8\pi k L}}
\, \sin (k L - \pi/4)
+ ...
\end{eqnarray}

	Equation (\ref{f6bis}) is the exact contribution of a family in an
integrable polygon. Therefore there are non-diffractive corrections to the
leading order (\ref{f2bis}) of the trace formula which are of same order as
(\ref{f6}). To ${\cal D}_{reg}$ in (\ref{f6}) one simply adds a factor ${\cal
O}({\cal A}/L^2)$ (see the last term of the r.h.s. of (\ref{f6bis})). In all
our numerical checks this correction appeared to be negligible. By comparing
the diffractive corrections with (\ref{f6bis}) one can note that~: (i) the
first diffractive term (\ref{f4}) is the leading $\sqrt{\hbar}$ correction in
the trace formula and (ii) for long orbits (or large repetition number) the
term ${\cal A}/L^2$ in (\ref{f6bis}) will be dominated by ${\cal D}_{reg}$ in
(\ref{f6}).

\

	$\bullet$ The determination of the contribution of the next iterates
of a primitive family of periodic orbits with a diffractive boundary
is patterned on the above derivation. We just state here the results. 

	The main contribution is the generic semi-classical one, 
of the form (\ref{f2bis}). The first correction is of a type similar to the
contribution to the trace formula of an isolated periodic orbit~:

\begin{equation}\label{f7}
\rho(E )\leftarrow
- \frac{\ell}{\pi k} \, C_n \cos (k L) \; ,\end{equation}

\noindent where $\ell$ is the primitive length, $n$ is the repetition number,
$L=n\,\ell$ and $C_n$ is a dimensionless parameter given by the formula
$C_n=(1/8\pi)\sum_{q=1}^{n-1}[q(n-q)]^{-1/2}$. We show how to compute it in
some special cases in \ref{integrales} and in general in \ref{valeurcn}.
Its first values are $C_1=0$, $C_2=1/(8\,\pi)$,
$C_3=1/(4\pi\sqrt{2})$ ... and it has the limiting value $C_\infty=1/8$.

	The next correction to (\ref{f7}) is of the form (\ref{f6}). This is proven
in special cases in \ref{integrales} (Eqs. (\ref{b10},\ref{b11})) and in
general in \ref{eugene}.

\

\begin{figure}[thb]
\begin{center}
\includegraphics*[height=6cm,bbllx=28pt, bblly=90pt, bburx=525pt,
bbury=555pt]{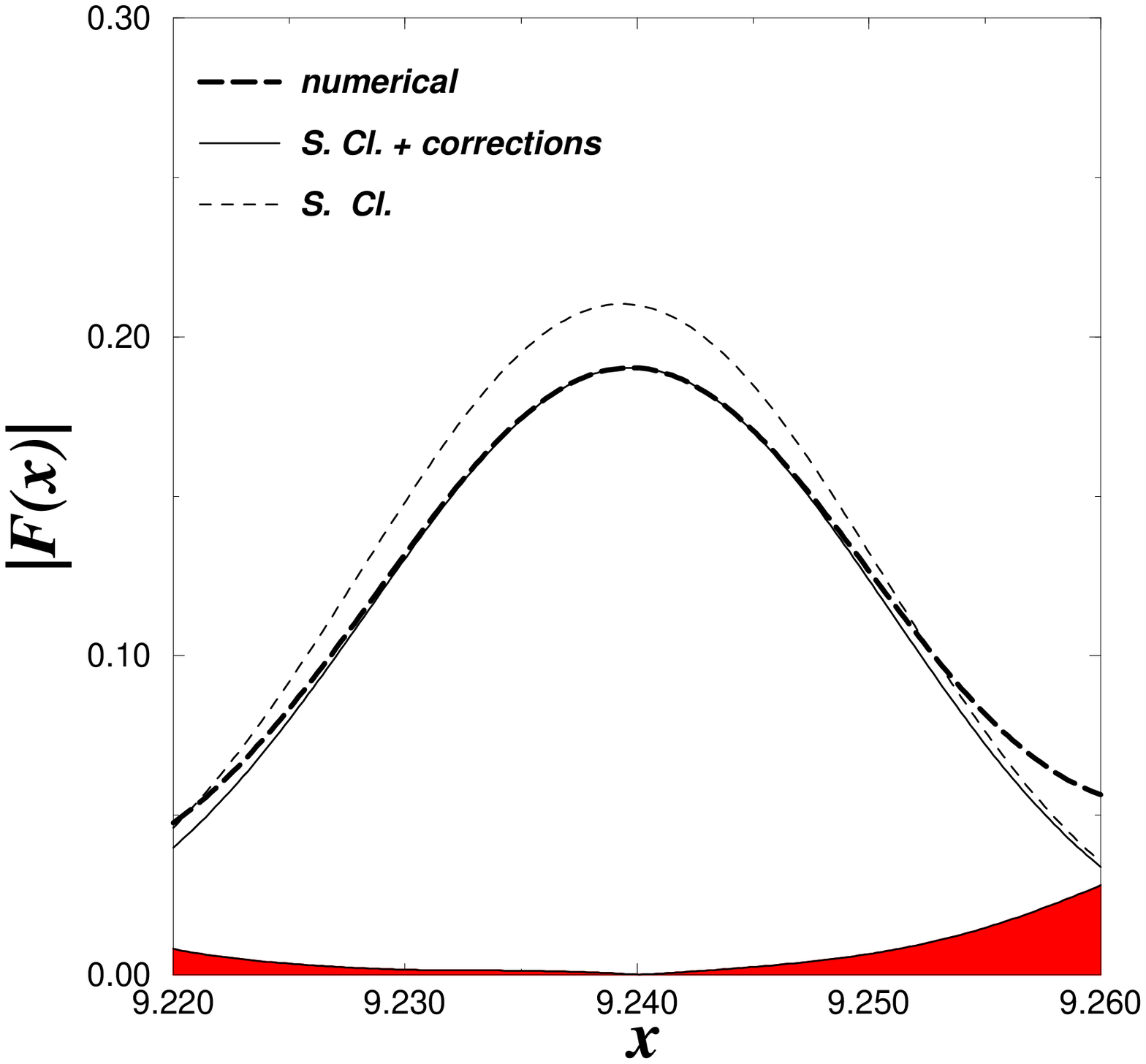}
\includegraphics*[height=6cm,bbllx=28pt, bblly=90pt, bburx=530pt,
bbury=555pt]{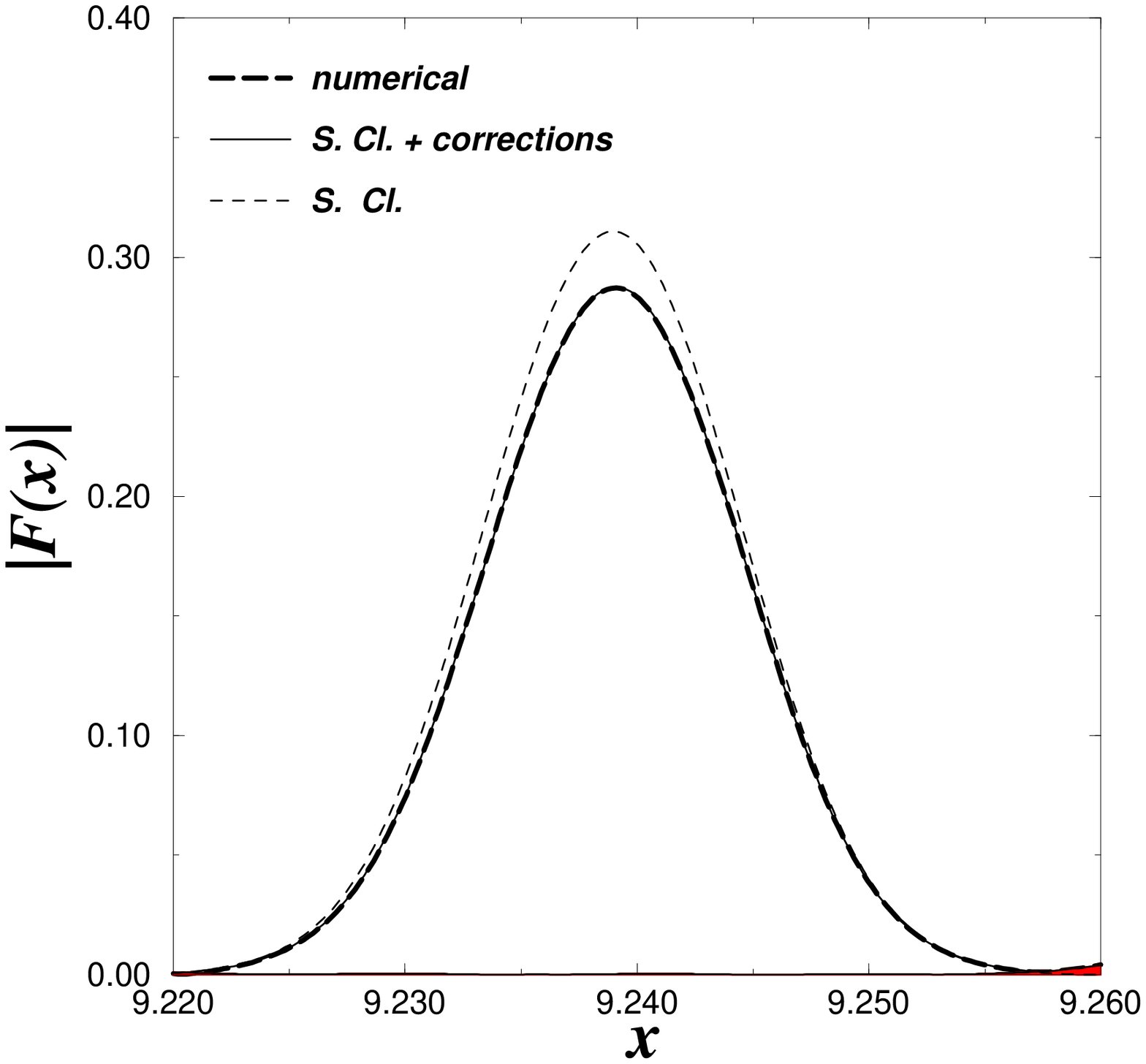}
\end{center}
\caption{\small Same as Fig.~\ref{fig1} for the fifth iterate of the family
shown in Figure~\ref{frontiere3} ($L\simeq 9.239$). The analytical result
corresponds to Eqs. (\ref{f2bis},\ref{f7},\ref{f6}) with $n=5$. The modulus of
the difference between the numerical and analytical $F(x)$ is represented by
the shaded area. It is important on the left plot, due orbits in vicinity of
the peak that are not taken into account. This overlapping of peaks can be
suppressed by increasing $k_{max}$ (in order to decrease the width of the
peaks in $|F(x)|$). This is done on the right plot which is drawn for
$k_{max}$ corresponding to the 20 000$^{th}$ level. Then, the largest
discrepancy with the numerical result is $4\times 10^{-4}$ whereas the error
when employing the usual semiclassical approach is 100 times larger.}
\label{fig3}
\end{figure}

	We have tested the excellent agreement of contributions
(\ref{f2bis},\ref{f7},\ref{f6}) with the numerical spectrum. We illustrate
this for the fifth iterate of the family shown in Fig.~\ref{frontiere3}. This
family is particular in the sense that one of its boundaries is formed by an
isolated orbit which has an extra bounce compared to the family~; it lies
along the lower edge of the triangles in Fig.~\ref{frontiere3}. The
contribution of such an isolated orbit has been known for some time
\cite{lau91,sie93} and is taken into account in the comparison with numerical
results displayed in Fig.~\ref{fig3}. The other boundary is a diffractive
orbit of the type we are interested in. Its contribution to the level density
is described by Eqs.~(\ref{f7},\ref{f6}).


\

	$\bullet$ The following diffractive corrections to the contributions
(\ref{f2bis},\ref{f7},\ref{f6}) to the $n^{th}$ iterate of a family correspond
to orbits having $n_k$ Kirchhoff diffractions and $n_g$ Keller ones, with
$n_k+n_g=n$. They yield corrections of order ${\cal O}(k^{-\frac{n_g+1}{2}})$
compared to the leading term (\ref{f2bis}). One can show that their
contribution to the level density is of the form~:

\begin{equation}\label{f8}
\rho(E)\leftarrow \frac{\ell}{2\pi k}
\left(\frac{D_{reg}}{\sqrt{8\pi k\ell}}\right)^{n_g}
B^{(n_g)}_n \, \cos(k L-3 n_g\pi/4 -\nu_d\pi/2)\; .\end{equation}

	The Maslov index $\nu_d$ in (\ref{f8}) is related to the index $\sigma$ of
the optical boundary considered (i.e. the one for which ${\cal
D}_{\sigma,\eta}$ in Eq.~(\ref{e3}) diverges) by $\exp \{-i\nu_{d}\pi/2\}
=\sigma^{n_g}$. $B^{(n_g)}_n$ is a dimensionless
coefficient. $B^{(1)}_n=n^{-1/2}$ (in agreement with (\ref{f6})), 
$B^{(2)}_n=\sum_{q=1}^{n-1}q^{-3/2}(n-q)^{-1/2}$ and the general form is~:

\begin{equation}\label{f9}
B^{(n_g)}_n = \sum_{\{q_i\} }\;
\prod_{i=1}^{n_g} \;(q_{i+1}-q_i)^{-3/2} \; ,\end{equation}

\noindent with the convention $q_{n_g+1}=n+q_1$. The sum is extended over all
possible sets of $n_g$ integers $\{q_i\}_{1\le i\le n_g}$ with $1\le q_1<
q_2<...<q_{n_g}\le n$.

\section{A diffractive orbit on the frontier of both a family and of 
the billiard}\label{deuxfront}

\begin{figure}[thb]
\begin{minipage}[c]{0.46\linewidth}
\begin{center}
\includegraphics*[width=8cm,bbllx=75pt, bblly=180pt, bburx=305pt,
bbury=290pt]{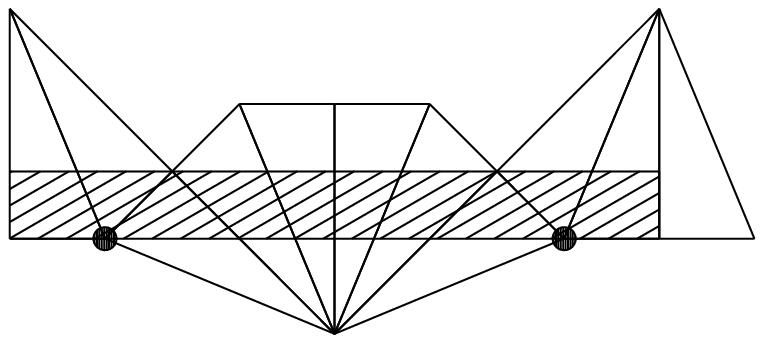}
\end{center}
\caption{\small Representation of a family of periodic orbits in the triangle
$(\pi/2,\pi/8,3\pi/8)$ by the technique of unfolding. The family has a length
$L=(4+2\sqrt{2})^{1/2}\simeq 2.613$. Its area is shaded and the diffractive
points on its boundary are marked with a black spot. The boundary of the
family party coincides with the frontier of the billiard.}
\label{deux_front1}
\end{minipage}\hfill
\begin{minipage}[c]{0.46\linewidth}
\begin{center}
\includegraphics*[width=8cm,bbllx=40pt, bblly=140pt, bburx=565pt,
bbury=450pt]{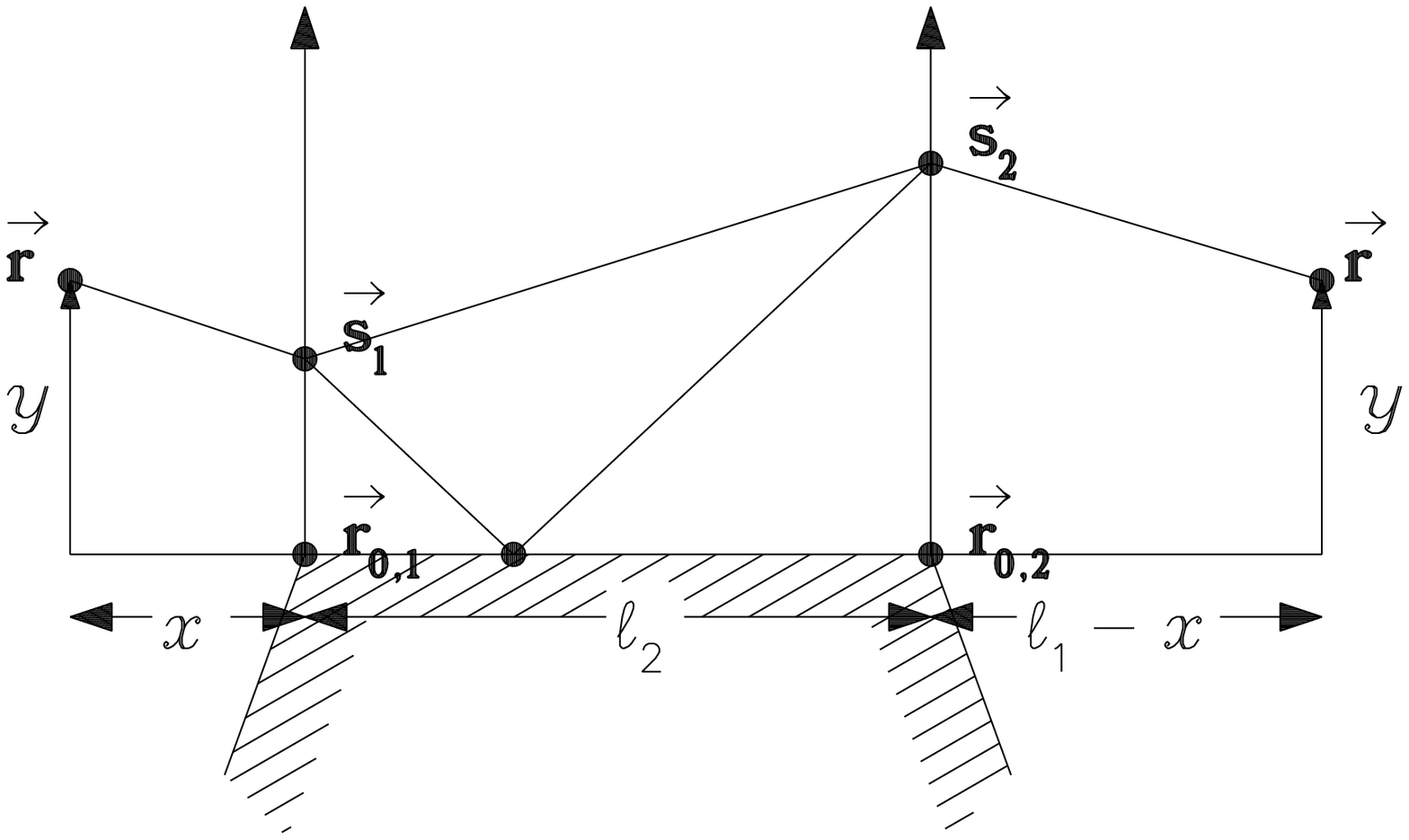}
\end{center}
\caption{\small Schematic representation of the different contributions to the
Green function near a diffractive periodic orbit on the boundary of a family.
The boundary of the family coincides with the one of the billiard along a
segment of length $\ell_2$. In this case, the main Kirchhoff term in
(\ref{f2}) contains two paths which are represented with solid lines of the
figure.}
\label{deux_front2}
\end{minipage}
\end{figure}

	In the previous Section we have studied the case that a diffractive
periodic orbit lies on an optical boundary corresponding to the frontier of
a family. There is a special configuration where such a diffractive orbit
lies on {\it two} optical boundaries. From Fig.~\ref{opt_bound} one sees
that two optical boundaries meet only on the edges defining the diffractive
corner (when $\t$ or $\t'=0$ or $\gamma$). Hence, in that case
part of the diffractive
trajectory crawls along the frontier of the billiard. This happens for
instance for the diffractive trajectory on the boundary of the family shown
on Fig.~\ref{deux_front1}.

	Although the diffractive periodic orbit considered bounds the first iterate
of a family, it is already doubly diffractive. For incorporating such a
configuration into the trace formula, one can still use Eqs. (\ref{e4}) and
(\ref{f2}), but the semi-classical Green function to be incorporated in that
formula has two contributions~: one from the ``direct'' path (we call this
path direct, but it may have bounces that are not materialized by the process
of unfolding) and one from a path bouncing on the frontier of the billiard
which is also the frontier of the family (see Fig. \ref{deux_front2}).

	In that figure one has represented a configuration where a point in
vicinity of the diffractive periodic orbit lies along the part of the
boundary of the family which does not coincide with one frontier of the
billiard (we denote by $\ell_1$ the length of this part, and by $\ell_2$ the
length of the part along a frontier of the billiard, $\ell_1+\ell_2=L$).
Then, the main Kirchhoff contribution to $G_2$ is of the form~:

\begin{eqnarray}\label{df1}
\frac{\sqrt{k}\,
\ep^{i k L+3i\pi/4}}{2\,(2\pi)^{3/2}\sqrt{x\,\ell_2(\ell_1-x)}}
\!\!\!\!\!\!\!\!\!& & 
\left[ \int_0^{+\infty}\!\!\!\!\!ds_1
\int_0^{+\infty}\!\!\!\!\!ds_2\,
\ep^{i\frac{k}{2}\left\{
\frac{(s_1-y)^2}{x}+
\frac{(s_2-s_1)^2}{\ell_2}
+\frac{(y-s_2)^2}{\ell_1-x} \right\} } \right.\nonumber \\
& & \nonumber\\
& -& \left. \int_0^{+\infty}\!\!\!\!\!ds_1
\int_0^{+\infty}\!\!\!\!\!ds_2\,
\ep^{i\frac{k}{2}\left\{
\frac{(s_1-y)^2}{x}+
\frac{(s_2+s_1)^2}{\ell_2}
+\frac{(y-s_2)^2}{\ell_1-x} \right\} }
\right]
\; .\end{eqnarray}

	The second contribution in (\ref{df1}) is obtained from the first one by the
method of images. It corresponds in Fig.~\ref{deux_front2} to the path going
from $\vec{s}_1$ to $\vec{s}_2$ with one bounce on the frontier of the
billiard. If the point $\ver$ of Fig. \ref{deux_front2} lies along the part of
the orbit coinciding with the frontier of the billiard, then the main
Kirchhoff contribution to $G_2$ is a sum of four terms (this is detailed in
\ref{integrales}, cf. Fig.~\ref{deux_front3}). We will not give the explicit
computation here (see \ref{integrales}), but after transverse integration
along $y$ the final result for the first diffractive correction to the
contribution of a family such as the one presented in Fig.~\ref{deux_front1}
is~:

\begin{equation}\label{df2}
\rho(E) \leftarrow
-\frac{1}{4\pi^2k}
\left(\sqrt{\ell_1\ell_2}+L \;\mbox{arctg}\,
\sqrt{\frac{\ds\ell_2}{\ds\ell_1}}\, \right)
\,\cos(k L)
\; .\end{equation}

\begin{figure}[thb]
\begin{center}
\includegraphics*[height=6cm,bbllx=28pt, bblly=90pt, bburx=525pt,
bbury=555pt]{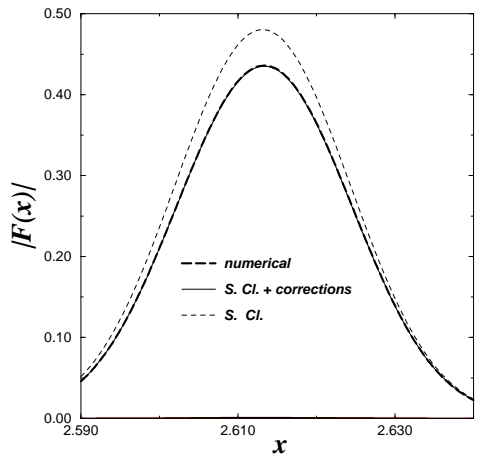}
\end{center}
\caption{\small Comparison between the numerical evaluation of $F(x)$ with the
result of Eqs.~(\ref{f2bis},\ref{df2}) for $x$ close to the length of the
family shown in Fig.~\ref{deux_front1}. The modulus of the difference is also
plotted on the figure, but cannot bee seen (it is lower than $10^{-3}$).}
\label{fig4}
\end{figure}

	The next diffractive corrections to (\ref{df2}) are of the order of
a doubly diffractive Keller correction, and we do not include them
in our description of the family. As seen in Fig.~\ref{fig4}, contributions
(\ref{f2bis})
and (\ref{df2}) already give an excellent description of the Fourier
transform of the spectrum in vicinity of the length of the family drawn on
Fig.~(\ref{deux_front1}).

\section{A diffractive orbit jumping from the boundary of a 
family to the boundary of an other one}\label{jump}

\begin{figure}[thb]
\begin{minipage}[c]{0.46\linewidth}
\begin{center}
\includegraphics*[width=8cm,bbllx=80pt, bblly=328pt, bburx=285pt,
bbury=390pt]{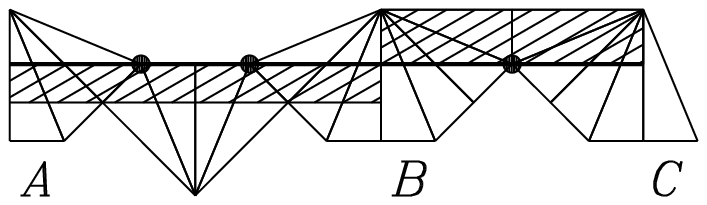}
\end{center}
\caption{\small Representation of a diffractive orbit that jumps from the
boundary of a family to the boundary of an other one. The area occupied by
each family is shaded. The first family transports triangle $A$ to position
$B$, and the second one transports triangle $B$ to position $C$. The
diffractive orbit is the thick solid line with its three diffractive points
marked by black dots. It has a length $L_d=(10+7\sqrt{2})^{1/2}\simeq 4.461$.}
\label{zigzag1}
\end{minipage}\hfill
\begin{minipage}[c]{0.46\linewidth}
\begin{center}
\includegraphics*[width=8cm,bbllx=40pt, bblly=105pt, bburx=565pt,
bbury=443pt]{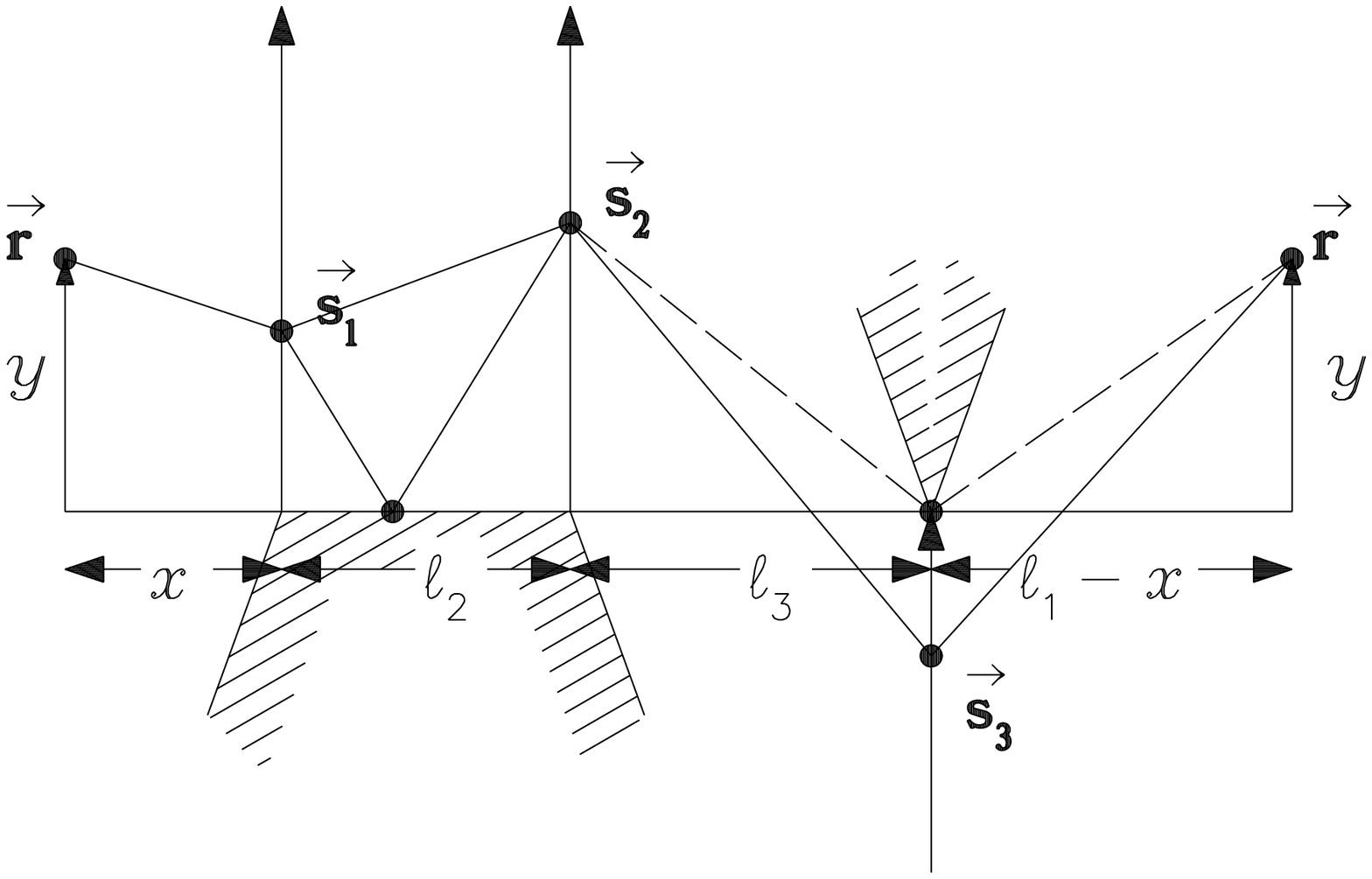}
\end{center}
\caption{\small Schematic representation of the contributions to the Green
function in vicinity of a diffractive orbit that jumps from the boundary of a
family to the boundary of another one (such as the one shown in
Fig.~\ref{zigzag1}). $\ell_1$ corresponds to the length of the family that
maps triangle $B$ onto triangle $C$ in Fig.~\ref{zigzag1} and $\ell_2+\ell_3$
to the length of the family that maps triangle $A$ onto $B$. For this last
family, $\ell_2$ is the length of the part of its boundary that lies along
the frontier of the billiard.}
\label{zigzag2}
\end{minipage}
\end{figure}

	In the billiard we consider, an interesting combination of orbits occurs. It
is formed by the gathering of two diffractive orbits, each being on the
boundary of a family, and where the total diffractive orbit is on the optical
boundary, although the two families have no overlap. An example of such a case
is given in Fig.~\ref{zigzag1}.

	As seen in Figure~\ref{zigzag1}, although the diffractive orbit lies
on the optical boundary, there is no allowed classical trajectory nearby.
This type of orbit might be a particularity of Veech billiards, but it is
nevertheless interesting to describe its contribution to the level
density. The schematic representation of the neighborhood of the orbit is
displayed in Fig.~\ref{zigzag2}.

	In the cases we have studied, the problem is complicated by the fact that
one of the families considered has a boundary that partially coincides with
the frontier of the billiard, i.e. is of the type studied in the previous
section. Hence, there are three relevant lengths along the orbit we consider~:
$\ell_1$ is the length of one of the families, $\ell_2+\ell_3$ is the length
of the second one, $\ell_2$ being the length that corresponds to the part of
the boundary of the second family lying along the frontier of the
billiard.

	The diffractive Green function to be considered here is of the the type
previously studied in Section \ref{deuxfront}, with an additional
diffractive bounce. 
Hence, one defines a Green function $G_3$ connected to $G_2$ in the same
manner as $G_2$ is connected to $G_1$ in Eq.~(\ref{f2}). Due to the
possible bounce along the frontier of the billiard that coincides with the
boundary of the family, the
semi-classical Green function to be incorporated in this formula has several
contributions. This is illustrated in Fig.~\ref{zigzag2} where
there are two possible paths for going from $\vec{s}_1$ to
$\vec{s}_2$ (such a contribution was already present in
Fig.~\ref{deux_front2}). 
If the initial point $\ver$ were lying near the frontier of the
billiard (next to the part of the family of length $\ell_2$),
one would have four different paths~: two for going from $\ver$ to
$\vec{s}_2$ and two for going from $\vec{s}_1$ to $\ver$. After transverse
integration of these four contributions (along the variable $y$),
one can verify that they lead to the
same contribution as the ones displayed in Fig.~\ref{zigzag2}, hence we will
only present here the computation in the simpler case shown in
Fig.~\ref{zigzag2}.

\

	For the configuration of Fig.~\ref{zigzag2}, the Kirchhoff term in
the expression of $G_3$ reads~: 
	
\begin{eqnarray}\label{zz1}
(2i k)^3\, D_3(x,\ell_1,\ell_2,\ell_3)
\!\!\!\! &\!\!\!\!  & \left[
\int_0^{+\infty}\!\!\!\!\!ds_1
\int_0^{+\infty}\!\!\!\!\!ds_2
\int_{-\infty}^{0}\!\!\!\!\!ds_3\,
\ep^{i\frac{k}{2}\left[
\frac{(s_1-y)^2}{x}+\frac{(s_2-s_1)^2}{\ell_2}+
\frac{(s_3-s_2)^2}{\ell_3}+\frac{(y-s_3)^2}{\ell_1-x}\right]}
\right.\nonumber\\
& & \nonumber\\
 & - \!\!\!\!& \left.
\int_0^{+\infty}\!\!\!\!\!ds_1
\int_0^{+\infty}\!\!\!\!\!ds_2
\int_{-\infty}^{0}\!\!\!\!\!ds_3\,
\ep^{i\frac{k}{2}\left[
\frac{(s_1-y)^2}{x}+\frac{(s_2+s_1)^2}{\ell_2}+
\frac{(s_3-s_2)^2}{\ell_3}+\frac{(y-s_3)^2}{\ell_1-x}\right]}
\right]
\; ,\end{eqnarray}

\noindent where the notation $D_n$ is defined in \ref{integrales}. In
(\ref{zz1}), the second term of the r.h.s. is obtained from the first one by
the method of image. It corresponds to a path going from $\vec{s}_1$ to
$\vec{s}_2$ with a specular bounce off the frontier of the billiard (cf.
Fig.~\ref{zigzag2}). Note that there is no classical path contributing to
(\ref{zz1})~: it is clear from Fig.~\ref{zigzag2} that there exits no
classical trajectory from $\ver$ to $\ver$. The transverse and longitudinal
integrations are done in a manner similar to the one exposed in
\ref{integrales} for the similar case of an orbit which boundary coincides
with the frontier of the billiard (cf. Section \ref{deuxfront}). The
contribution of the diffractive orbit to the level density is~:

\begin{eqnarray}\label{zz2}
\rho(E)\leftarrow -\frac{L_d}{8\pi^2 k}\,\cos (k L_d) \left\{
\,\mbox{arctg}\,\sqrt{\frac{\ds\ell_2}{\ds\ell_3+\ell_1}}\right.
\!\!\!\! & - & \!\!\!\!
\,\mbox{arctg}\,\sqrt{\frac{\ds\ell_3+\ell_2}{\ds\ell_1}}
+\,\mbox{arctg}\,\sqrt{\frac{\ds\ell_3}{\ds\ell_1+\ell_2}}
\nonumber\\
& & \nonumber\\
-\frac{\sqrt{\ds\ell_3(\ell_1+\ell_2)}}{\ds L_d}\!\!\!\!
& + & \!\!\!\!\left.
\frac{\sqrt{\ds\ell_2(\ell_3+\ell_1)}}{\ds L_d}
-\frac{\sqrt{\ds\ell_1(\ell_2+\ell_3)}}{\ds L_d}
\right\}
\; ,\end{eqnarray}

\noindent where $L_d=\ell_1+\ell_2+\ell_3$ is the length of the diffractive
orbit.

\

	In order to have a good description of the contribution of the orbit
we are considering here, one needs (as in Section \ref{front}) to
incorporate next 
order corrections, i.e. mixed Kirchhoff-Keller terms. This
corresponds in Fig.~\ref{zigzag2} to the path with one Keller bounce on the
apex which is not on the frontier of the billiard (dashed line) (Keller
bounces on the other apexes contribute to higher order). We do not detail
the computation here and just present the resulting correction to
(\ref{zz2}). It is of the form~:

\begin{equation}\label{zz3}
\rho(E)\leftarrow
\frac{L_d}{2\pi^2 k}\,\frac{\ds{\cal D}_{reg}}{\sqrt{\ds 8\pi k L_d}}
\,\mbox{arctg}\,\sqrt{\frac{\ds\ell_3\ell_1}{\ell_2 L_d}} \,
\cos(k L_d -\nu_d\pi/2 - 3\pi/4)
\; .\end{equation}

\begin{figure}[thb]
\begin{center}
\includegraphics*[height=6cm,bbllx=15pt, bblly=90pt, bburx=530pt,
bbury=555pt]{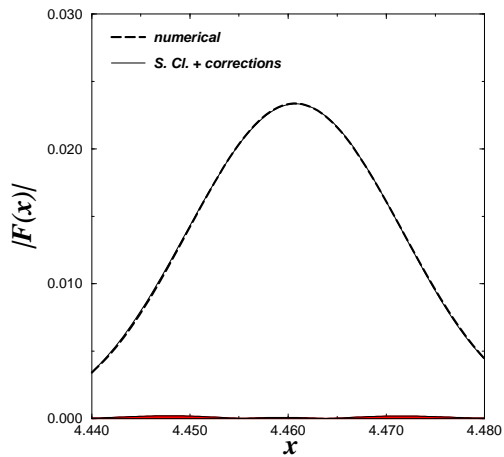}
\end{center}
\caption{\small Same as Fig.~\ref{fig1} for the orbits shown of
Fig.~\ref{zigzag1}. The solid line corresponds to the results of
Eqs.~(\ref{zz2},\ref{zz3}). The shaded area at the bottom corresponds to the
modulus of the difference between the numerical and the analytical result
(which is inferior to $2\times 10^{-4}$). Note that the standard
semi-classical approach completely misses this peak in $|F(x)|$.}
\label{fig5}
\end{figure}

	As one can see in Fig.~\ref{fig5}, Eqs.~(\ref{zz2},\ref{zz3}) give an
excellent account of the contribution to the level density of the orbit shown
on Fig.~\ref{zigzag1}.

\section{A diffractive orbit bouncing between the upper and lower
boun\-dary of a family}\label{updown}

\begin{figure}[thb]
\begin{minipage}[c]{0.46\linewidth}
\begin{center}
\includegraphics*[width=8cm,bbllx=75pt, bblly=215pt, bburx=330pt,
bbury=370pt]{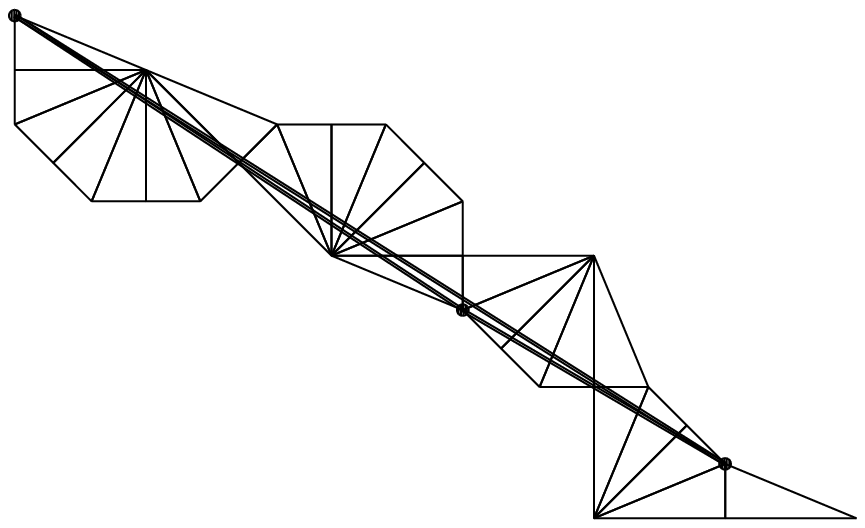}
\end{center}
\caption{\small A diffractive orbit on the boundary of a family (straight line
between two black dots). The family is not represented, but it connects the
upper left triangle to the lower right one. The other orbit shown is typical
of those studied in this section. It is a doubly diffractive orbit close to
the family, it is represented by the segmented line between three black dots.}
\label{updown1}
\end{minipage}\hfill
\begin{minipage}[c]{0.46\linewidth}
\begin{center}
\includegraphics*[width=8cm,bbllx=54pt, bblly=150pt, bburx=551pt,
bbury=470pt]{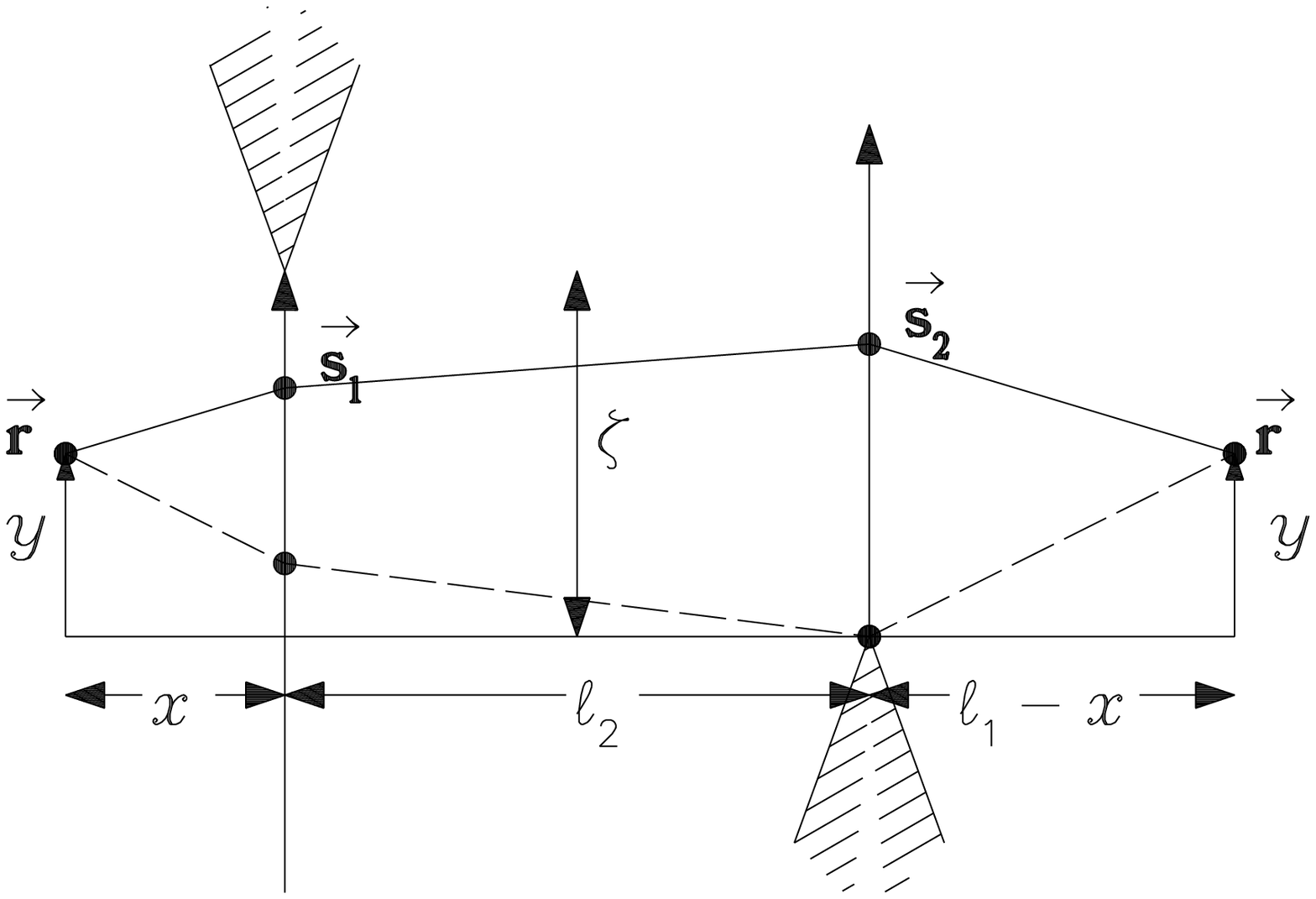}
\end{center}
\caption{\small Graphical representation of different diffractive
contributions for a family limited by two opposite wedges. The leading term in
(\ref{f2}) corresponds to the solid line. The dashed path is one of the next
order corrections, involving one Keller bounce (with a coefficient ${\cal
D}_{reg}$) on one of the apexes. $\zeta$ is the distance between the two
apexes measured transverse to the direction of the family.}
\label{updown2}
\end{minipage}
\end{figure}

	Up to now we have only considered diffractive orbits which were lying
exactly on the optical boun\-da\-ry. Other types of diffractive orbit occur
which do not stand right on the optical boundary, but close enough to prevent
their description by the geometrical theory of diffraction. Such an orbit is
represented in Fig.~\ref{updown1}.

	On this Figure, the upper left triangle and the lower right one are
connected by a family. For legibility we do not represent it and its
area. We only represent the diffractive orbit on its boundary (the
straight line between two black dots). This orbit is singly diffractive and
its contribution corrects the one of the family as in Eq.~(\ref{f1}). There is
a diffractive orbit nearby, not exactly on the optical boundary, but very
close to being part of the family~: it starts and ends at the same point as
the diffractive orbit on the boundary of the family, but it has an extra
diffractive bounce in between (see Figure \ref{updown1}). This is the type of
orbits we aim to describe in this section. Its diffraction coefficient does
not exactly diverge, but one of the contributions ${\cal D}_{\sigma,\eta}$ of
(\ref{e3}) is large and does not allow a proper description of the diffractive
Green function by means of Eq.~(\ref{e2}). For simplicity we will denote this
part as the ``divergent part'' (the remaining being the ``regular part'').

	The configuration we just described is of the type represented schematically
on Fig.~\ref{updown2}. The projection of the upper diffractive corner onto the
family separates it in two parts of lengths $\ell_1$ and $\ell_2$
($\ell_1+\ell_2=L$). Typically, the upper wedge represented in that figure is
the upper boundary of the family. In that case, if $\zeta$ denotes the
distance from the upper wedge to the lower extend of the family, the area
occupied by the family is simply ${\cal A}=\zeta\,L$ ($L$ being the length of
the family, see Figure~\ref{updown2}).

	In this configuration the leading term in the
Green function is obtained from the explicit expression of $G_2$ and reads~:

\begin{equation}\label{ud1}
G_2(\ver,\ver,E) \approx
\frac{\sqrt{\ds k}\, \ep^{\ds i k L+3i\pi/4}}
{2\, (2\pi)^{3/2}\sqrt{\ds x \ell_2 (\ell_1-x)}}\;
\int_{-\infty}^{\zeta}\!\!\!\!\!ds_1\int_0^{+\infty}\!\!\!\!\!ds_2\;
\ep^{i\frac{k}{2}\left\{
\frac{(s_1-y)^2}{x}+\frac{(s_2-s_1)^2}{\ell_2}+\frac{(y-s_2)^2}{\ell_1-x}
\right\} }\; .
\end{equation}

	The above expression integrated transversely (along $y$) and
longitudinally (along $x$) gives the contribution of the family and of its
corrections to the level density. The result reads (the relevant integrals
are given in \ref{integrales})

\begin{eqnarray}\label{ud2}
-\frac{1}{\pi}\,\mbox{Im}\,\int d^2r \, G_2(\ver,\ver,E) & \approx & 
{\ds {\cal A}\over\ds 2 \pi}
\frac{1}{\sqrt{\ds 2\pi k L}}
\, \cos (k L -\pi/4)\nonumber \\
 & & \nonumber\\
 & + \frac{\sqrt{\ds\ell_1\ell_2}}{\ds 4\pi^2\,k} &
\left[ \cos(k L_d) - 
2\, \sqrt{\pi k \Delta} \; \mbox{Re}\,
\left\{ \ep^{i k L_d-i\pi/4} K\left(\sqrt{k\Delta }\right) \right\}
\right]
\; .\end{eqnarray}

	In (\ref{ud2}), $K$ is the modified Fresnel function defined in \ref{unif}.
We have denoted by $L_d$ the length of the doubly diffractive orbit going from
the upper corner to the lower one ($L_d=\sqrt{\ell_1^2+\zeta^2} +
\sqrt{\ell_2^2+\zeta^2}$), by $\Delta$ the length difference $L_d-L$,
and have made the approximation $\Delta\simeq \zeta^2 L/(2\ell_1\ell_2)$. The
first term of the r.h.s. of (\ref{ud2}) is the usual contribution of a family.
The second is a diffractive correction.

\

	Two remarks are in order here~:

\

$\bullet$ It may happen that the upper corner of Fig.~\ref{updown2} does not
provide the upper boundary of the family because the family meets an other
non-diffractive boundary between the two diffractive corners. This is the case
presented in Fig.~\ref{updown1}~; the family does not occupy all the width
between the two diffractive corners~: it meets first a non-diffractive $\pi/2$
corner. In this case formula (\ref{ud2}) remains valid, but ${\cal
A}=\zeta\,L/2$.

\

$\bullet$ Secondly, it is interesting to check what is the behavior of
Eq.~(\ref{ud2}) when the two wedges are far apart, i.e. in the limit
$\sqrt{k\Delta}\gg 1$. By using the asymptotic expansion (\ref{a3}) of the
modified Fresnel function one obtains~:

\begin{eqnarray}\label{ud3}
-\frac{1}{\pi}\,\mbox{Im}\,\int d^2r \, G_2(\ver,\ver,E)
& \underset{\sqrt{k\Delta}\gg 1}{\approx} &
{\ds {\cal A}\over\ds 2 \pi}
\frac{1}{\sqrt{\ds 2\pi k L}}
\, \cos (k L -\pi/4)\nonumber\\
& & \nonumber\\
& & +
\frac{L_d}{2\pi k}\left(\frac{2\ell_1\ell_2}{L_d\Delta}\right)
\frac{1}{8\pi k\sqrt{\ell_1\ell_2}} \cos(k L_d-3\pi/2)\; .
\end{eqnarray}

	In (\ref{ud3}), to the usual contribution of a family is added a term which
can be matched with a contribution such as (\ref{e3bis}) with two diffractive
bounces, provided some approximations are made. The term
$2\ell_1\ell_2/(L_d\Delta)$ stands were one would expect a product of two
coefficients ${\cal D}$. Indeed, one can show that this term corresponds to
the product of the two divergent parts ${\cal D}_{\sigma,\eta}$ near the
optical boundary. But it is not of the form (\ref{e3}) which is the only
acceptable in the limit where Eq.~(\ref{ud3}) has been written (i.e. far from
the optical boundary). This is a well known drawback of Kirchhoff's
approximation already discussed in Section \ref{vob}. It can be cured
relatively easily~: if the optical boundary close to the diffractive orbit is
characterized by the indices $\sigma$ and $\eta$, one has to multiply the
second terms of the r.h.s. of (\ref{ud2}) by the factor
$(|a^d_{\sigma,\eta}|\,{\cal D}^d_{\sigma,\eta})^2/2$ and to express $\Delta$
as $|a^d_{\sigma,\eta}|^2 \ell_1\ell_2/L_d$ (instead of $L_d-L$). The term
$a_{\sigma,\eta}$ appearing in these expressions is defined in Eq.~(\ref{a4}).
The upper index $d$ is meant to remind that $a_{\sigma,\eta}$ and ${\cal
D}_{\sigma,\eta}$ have to bee evaluated on the diffractive periodic orbit of
length $L_d$. This procedure allows recovery of the correct limit in
(\ref{ud3}). Moreover it does not affect (\ref{ud2}) when the diffractive orbit
is close to the family (i.e. in the limit $\sqrt{k\Delta}\ll 1$) since in this
limit $(|a^d_{\sigma,\eta}|\,{\cal D}^d_{\sigma,\eta})^2/2\simeq 1$ and
$\Delta\simeq |a^d_{\sigma,\eta}|^2 \ell_1\ell_2/L_d$.

\

\begin{figure}[thb]
\begin{center}
\includegraphics*[height=6cm,bbllx=28pt, bblly=90pt, bburx=525pt,
bbury=555pt]{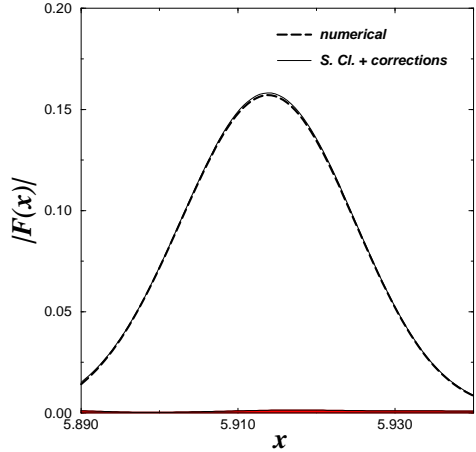}
\end{center}
\caption{\small Same as Fig.~\ref{fig1} for the orbits shown on
Fig.~\ref{updown1}. The solid line corresponds to Eqs.~(\ref{ud2},\ref{ud4}).
Here one has $L=(1+\sqrt{2})\sqrt{6} \simeq 5.9136$ and $L_d=
(10+3\sqrt{2})^{1/2}+(6-\sqrt{2})^{1/2}\simeq 5.9154$. The shaded area hardly
seen at the bottom of the plot corresponds to the modulus of the difference
between the numerical and the analytical result (which is inferior to
$1.4\times 10^{-3}$). The pure semi-classical estimate (Eq.~(\ref{f2bis}))
gives an error of $4.6\times 10^{-2}$. Taking only into account the
diffractive periodic orbit standing exactly on the boundary of the family (as
in Eq. (\ref{f1})) gives an even larger error ($5.3\times 10^{-2}$).}
\label{fig6}
\end{figure}

	Eq. (\ref{ud2}) is not the final contribution from the configuration
represented in Figure~\ref{updown2}. This is clear from (\ref{ud3})~: far from
the optical boundary the asymptotic evaluation of (\ref{ud2}) only allows to
recover the divergent part of the diffraction coefficient. Hence one has to
include other terms, of mixed type Keller-Kirchhoff, as already encountered in
Sections \ref{front} and \ref{jump}. These involve the regular part ${\cal
D}_{reg}$ of the diffraction coefficients. We must be careful though, that
one has now two different diffraction coefficients~: one for the orbit along
the boundary of the family (we denote it by ${\cal D}_{reg}^f$) and an other
one for the orbit bouncing from the lower wedge to the upper one (we denote it
by ${\cal D}_{reg}^d$). The remaining contribution to $\rho(E)$ of the
configuration considered in this section is (we do not detail the
derivation)~:

\begin{eqnarray}\label{ud4}
\rho(E)& \leftarrow & 
\frac{L}{2\pi k}
\frac{\ds {\cal D}^f_{reg}}{\sqrt{\ds 8\pi k L}}
\, \cos (k L -\nu_d\pi/2 - 3 \pi/4)\nonumber\\
& & \nonumber\\
& + & \frac{L}{\pi k}
\frac{\ds {\cal D}^d_{reg}}{\sqrt{\ds 8\pi k L}}
\,\mbox{Re}\,\left\{
\ep^{i k L_d+i\pi/4} K\left(\sqrt{k\Delta}\right) \right\}
\nonumber\\
& & \nonumber\\
& + & \frac{L_d}{2\pi k}
\frac{\left({\cal D}_{reg}^d\right)^2}{8\pi k\sqrt{\ell_1\ell_2}}
\cos(k L_d - 3 \pi/2)
\; .\end{eqnarray}

	The term involving a modified Fresnel function in the above expression can
be made uniform by a procedure similar to the one devised for Eq.~(\ref{ud2}).
Note also that we have added in (\ref{ud4}) a doubly diffractive term of
purely Keller type (last term of the r.h.s.). It is a small correction and
such terms were neglected in the previous sections. We kept it here for
consistency because far from the optical boundary, it is of same order as the
second term of the r.h.s. of (\ref{ud2}).

	The agreement with the numerical spectrum is here also excellent, as shown
by Figure~\ref{fig6}. Note that the geometrical theory of diffraction --
although yielding a non-divergent result -- is completely inadequate in this
case. It amounts here to treating the isolated diffractive orbit as truly
isolated from the family~: hence to describing the family of periodic orbits
in the usual way (i.e. using Eq.~(\ref{f2bis})) and including a correction of
type (\ref{e3bis}) describing the contribution of the doubly diffractive orbit
bouncing between the upper and lower boundary of the family. This procedure
gives an error of $9.4\times 10^{-2}$ in Fig.~\ref{fig6}.

\section{A diffractive orbit near an isolated one}\label{section7}

	In this Section we will study, as in the previous one, a diffractive
orbit standing not exactly on the optical boundary, but close to an allowed
`periodic orbit. Here we consider the case that the nearby orbit is
an isolated one (and not part of a family as in the previous section).
Such a configuration has already been studied in Ref.~\cite{sie97}, and we
will here re-derive the result in a simpler manner (but with less generality).

\begin{figure}[thb]
\begin{minipage}[c]{0.46\linewidth}
\begin{center}
\includegraphics*[height=4cm,bbllx=77pt, bblly=131pt, bburx=300pt,
bbury=306pt]{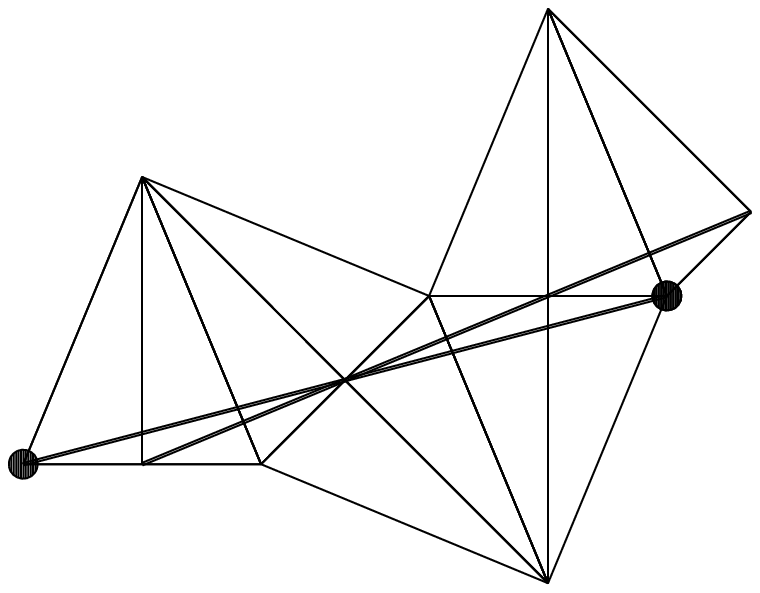}
\end{center}
\caption{\small A diffractive orbit (straight line between two black points)
near an isolated orbit (straight line connecting two corners with opening
angle $\pi/2$).}\label{isolee1}
\end{minipage}\hfill
\begin{minipage}[c]{0.46\linewidth}
\begin{center}
\includegraphics*[height=5cm,bbllx=165pt, bblly=114pt, bburx=570pt,
bbury=500pt]{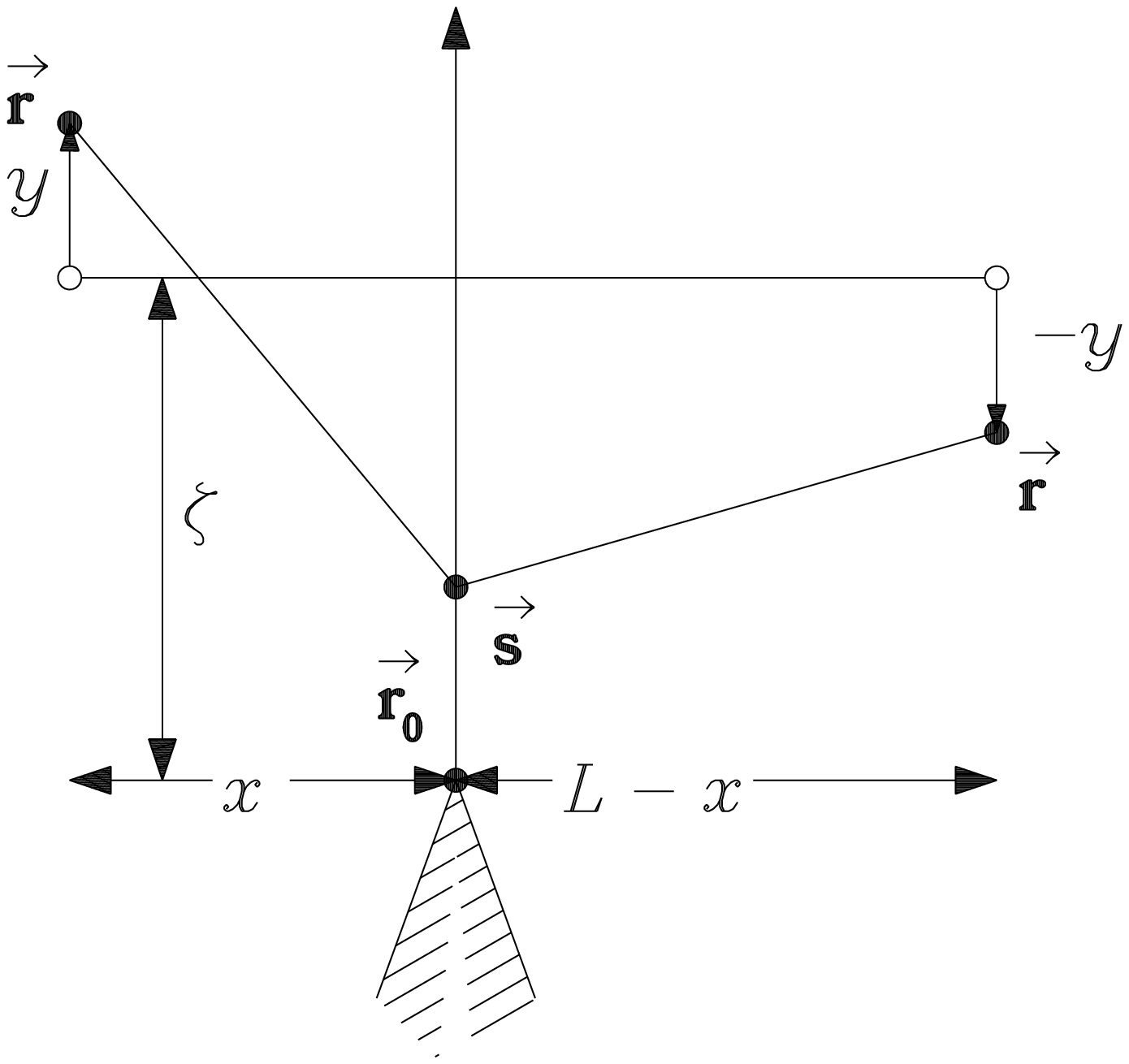}
\end{center}
\caption{\small In this figure the isolated periodic orbit connects the two
open circles. A nearby path from $\ver$ to $\ver$ is represented. There is an
inversion along the periodic orbit for the transverse coordinate $y$ (see the
text). The diffractive periodic orbit goes from one open circle to the apex at
$\vr0$ and to the other open circle. It has a length $L_d\simeq
L+2\zeta^2/L$.}\label{isolee2}
\end{minipage}
\end{figure}

	A typical occurrence of the situation we are interested in is shown
on Fig.~\ref{isolee1}. The isolated orbit we consider is the third iterate
of the shortest classical periodic orbit of the system. It has a length
$L=3/\sqrt{2}\simeq 2.121$. The nearby singly diffractive orbit
has a length $L_d=(6-\sqrt{2})^{1/2}\simeq 2.141$.

\

	The different contributions to the Green function $G_1$ are illustrated in
Fig.~\ref{isolee2}. Note that for the phase-space coordinate transverse to the
direction of an orbit, a bounce on a straight segment leads to an inversion.
Hence, in a polygonal enclosure, the transverse mapping near a periodic orbit
is either an inversion (for an odd number of bounces) and the orbit is then
isolated, or the identity (for an even number of bounces) and the orbit is
then part of a family. This is the reason for the inversion on
Fig.~\ref{isolee2} of point $\ver$ with respect to the axis of the isolated
periodic orbit after the process of unfolding. In the Figure, $\zeta$ is the
distance from the diffractive apex to the periodic orbit.

	In the case of interest here, the Kirchhoff part of the 
total Green function $G_1$ is (from Eq.~(\ref{e4}) and Fig.~\ref{isolee2})~:

\begin{equation}\label{i1}
- \frac{\ep^{\ds i k L-i\nu\pi/2}}{4\pi\sqrt{x(L-x)}} \,
\int_0^{+\infty}\!\!\!\!\!ds_1\,
\exp\left\{ i\frac{k}{2}\left[
\frac{(y+\zeta-s_1)^2}{x}+\frac{(\zeta-y-s_1)^2}{L-x}
\right]\right\}
\; ,
\end{equation}

\noindent where $\nu$ is the Maslov index of the isolated orbit
($\exp\{i\nu\pi/2\}=-1$). If $\nu_d$ is the one of the diffractive orbit and
if $\sigma$ characterizes the nearby optical boundary, one has
$\exp\{i\nu_d\pi/2\}=\sigma \exp\{i\nu\pi/2\}$. Once $G_{0}$ has been
removed, the above expression yields -- after transverse and
longitudinal integration -- the main contribution of the diffractive orbit to
the level density. There is also a corrective term containing the regular part
of the diffraction coefficient. Altogether one obtains the following
contribution~:

\begin{equation}\label{i2}
\rho(E) \leftarrow \frac{-L}{4\pi k} \,\mbox{Re}\,
\left\{ \ep^{i k L_d-i\nu\pi/2} K\left( \sqrt{k\Delta}\right) \right\} +
\frac{L_d}{2 \pi k} 
\frac{\ds {\cal D}_{reg}}{\sqrt{\ds 8\pi k L_d}}
\, \cos (k L_d -\nu_d\pi/2-3 \pi/4) \; ,
\end{equation}

\noindent where $\Delta=L_d-L\simeq 2\zeta^2/L$. As in the previous section we
have used a representation of the Green function based on Kirchhoff's
approximation which does not yield a uniform formula~: Eq.~(\ref{i2}) does not
permit recovery of the result of the geometrical theory of diffraction far
from the optical boundary, i.e. when the isolated and the diffractive orbit
are far apart. As in Section \ref{updown}, one can easily remedy this
deficiency. If the optical boundary to which the diffractive orbit is close is
characterized by the indices $\sigma$ and $\eta$, one multiplies the first
term of the r.h.s. of (\ref{i2}) by $\sigma (L_d/L) |a_{\sigma,\eta}|\, {\cal
D}_{\sigma,\eta}/\sqrt{2}$ and replaces $\sqrt{\Delta}$ in the argument of the
modified Fresnel function by $|a_{\sigma,\eta}|\,\sqrt{L_d}/2$. This procedure
does not affect (\ref{i2}) in the limit that the diffractive and isolated
periodic orbits are close and it allows recovery of the result of the
geometrical theory of diffraction when these two orbits are well separated.

\begin{figure}[thb]
\begin{center}
\includegraphics*[height=6cm,bbllx=18pt, bblly=90pt, bburx=525pt,
bbury=555pt]{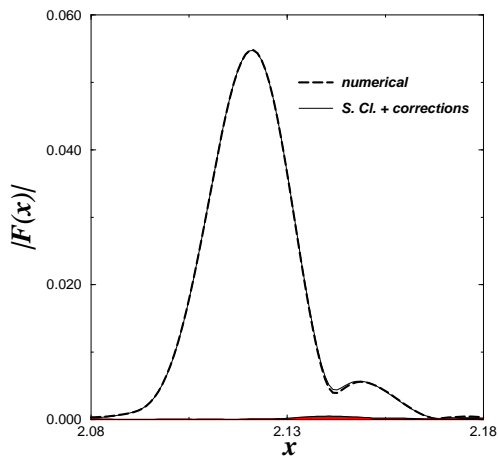}
\end{center}
\caption{\small Same as Fig.~\ref{fig1} for the orbits shown on
Fig.~\ref{isolee1}. The solid line corresponds to the contribution of the
isolated diffractive orbit (which has a length $L=3/\sqrt{2}\simeq 2.121$)
plus the contribution of the nearby diffractive orbit
($L_d=(6-\sqrt{2})^{1/2}\simeq 2.141$). The shaded area at the bottom of the
plot corresponds to the modulus of the difference between the numerical and
the analytical result (which is inferior to $5\times 10^{-4}$). We have used
here the uniformized version of Eq.~(\ref{i2}) (see the text), the use of the
plain formula gives twice a larger discrepancy with numerical datas.}
\label{fig7}
\end{figure}

	The comparison of formula (\ref{i2}) with the numerical result is very good,
as shown in Fig.~\ref{fig7}. Note that the geometrical theory of diffraction
is not totally inadequate here (as it was in the previous section). It gives
an error only four times larger than our approach. The reason is that the
classical and diffractive orbits considered here are not very close to each
other. Of course, the distance between two orbits must be measured relatively
to the wave length. As a result the accuracy of the geometrical theory of
diffraction depends on the window of the spectrum chosen for
evaluating $F(x)$. For instance, evaluating $F(x)$ keeping only the first 500
levels (instead of the first 5000 levels as in Fig.~\ref{fig7}) gives for the
geometrical theory of diffraction an error 10 times larger than our approach.

\section{A rectangular billiard with a flux line}\label{section8}

\begin{figure}[thb]
\begin{center}
\includegraphics*[height=6cm,bbllx=50pt, bblly=140pt, bburx=570pt,
bbury=470pt]{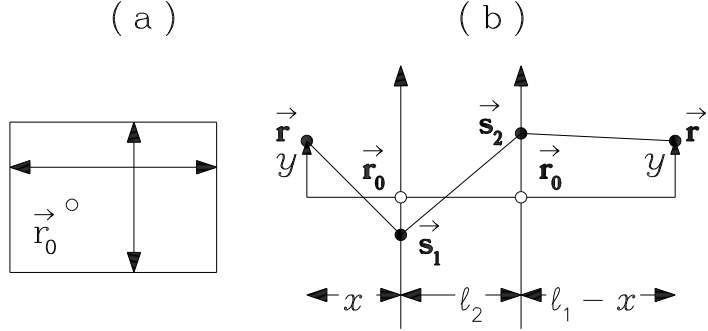}
\end{center}
\caption{\small Part (a)~: the rectangular billiard and the two bouncing ball
orbits (the vertical and the horizontal one). The flux line is located at
point $\vr0$ marked by a white circle. Part (b)~: representation of the
typical contribution to the Green function (\ref{e5}) in vicinity of the
doubly diffractive orbit belonging to one of the bouncing ball families. The
path going from $\vec{r}$, to $\vec{s}_1$, $\vec{s}_2$ and back to $\vec{r}$
represented in the figure accumulates a phase $\exp(-2 i\pi\alpha)$) (see the
text).}
\label{AB_rect}
\end{figure}

	In this Section we depart from the previous examples which treat corner
diffraction, and we consider instead diffraction by a flux line. We consider a
rectangular billiard (with sides of length $a$ and $b$) with a flux line
located at point $\vr0$ inside the billiard (cf.
Fig.~\ref{AB_rect}a).

	We will not restart here a detailed study of a large number of different
cases of diffraction in the system (as was done in Secs.~\ref{section2} to
\ref{section7} for a triangular billiard). First, because Aharonov-Bohm
diffraction is in a sense simpler than corner diffraction and leads to fewer
exceptional cases~; second because we chose this example merely to illustrate
the flexibility of Kirchhoff's approach devises in Sec. \ref{section2}. We
will show that Eq.~(\ref{e5}) permits to tackle the problem of multiple
forward Aharonov-Bohm scattering.

	This problem is encountered for instance when evaluating the contribution to
the trace formula of the two families drawn in Fig. \ref{AB_rect}a. For each
of these families the periodic orbit that encounters the point $\vr0$ twice
on its way gives a doubly diffractive contribution. The schematic contribution
to (\ref{e5}) for a nearby closed path is illustrate in Fig.~\ref{AB_rect}b.
In this Figure there is a reflection on the frontier of the billiard between
the two flux lines and this has the effect of changing the sign of $\alpha$ on
the second flux line (equivalently one could keep the same $\alpha$ and change
the orientation of the axis $(\vr0,s_2)$). From Eq.~(\ref{e5}) the diffractive
Green function of the problem is written~:

\begin{eqnarray}\label{ab1}
G_{2d}(\vec{r},\vec{r},E) & = & 
\frac{\sqrt{k}\, \ep^{i(k\ell-\pi/2)}}{2\pi\sqrt{8\pi x\ell_2(\ell_1-x)}}
\\ \nonumber
 & & \\ \nonumber
& &
\int_{-\infty}^{+\infty}\!
\int_{-\infty}^{+\infty}\!\!\!\!\!ds_1 ds_2 \;
\ep^{\frac{i k}{2}\left\{
\frac{(y-s_1)^2}{x}+\frac{(s_2-s_1)^2}{\ell_2}
+\frac{(y-s_2)^2}{\ell_1-x}\right\}}
\left[\ep^{i\pi\alpha\{\mbox{\small sgn}\,(s_1)-
\,\mbox{\small sgn}\,(s_2)\} }-1\right]
\; ,\end{eqnarray}

\noindent where $\ell$ is the length of the periodic orbit. The flux line
(encountered twice) separates the orbit in three parts having lengths denoted
by $x$, $\ell_2$ and $\ell_1-x$ in (\ref{ab1}) and Fig.~\ref{AB_rect}b
($\ell_1+\ell_2=\ell$). Transverse integration yields~:

\begin{equation}\label{ab2}
\int_{-\infty}^{+\infty}\!\!\!\!\!dy \; G_{2d}(\vec{r},\vec{r},E) =
-\frac{\sqrt{\ell_1\ell_2}}{2\pi k \ell}\;
\ep^{\ds i k\ell+i\pi/2}\left\{\cos(2\pi\alpha)-1\right\}
\; ,\end{equation}

\noindent and this gives a contribution to the level density~:

\begin{equation}\label{ab3}
\rho(E)\leftarrow - 
\frac{\sqrt{\ell_1\ell_2}}{\pi^2 k}
\sin^2(\pi\alpha)\cos(k\ell)\; .
\end{equation}

\begin{figure}[thb]
\begin{center}
\includegraphics*[height=6cm]{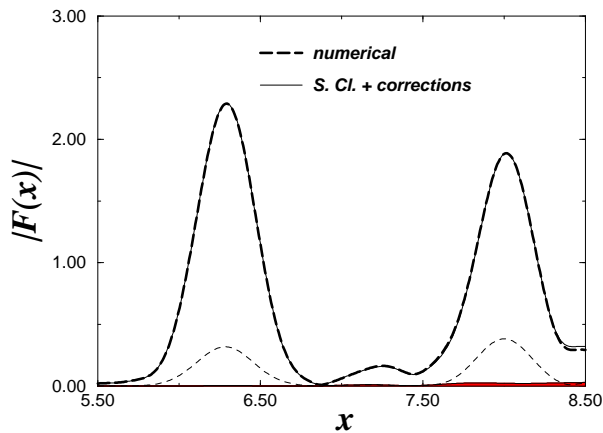}
\end{center}
\caption{\small Comparison of the numerical $|F(x)|$ (dashed line) with the
result from Eqs.~(\ref{f2bis},\ref{ab3}) (solid line). The large peaks
correspond to the lengths of the families shown in Fig.~\ref{AB_rect}a. In our
computations we have taken the sides of the rectangle to be $a=4$ and $b=\pi$
and the bouncing ball
families have then a length $\ell=8$ and $2\pi$. The shaded line
barely seen at the bottom is the modulus of the difference between the
numerical and analytical $F(x)$. We also show as a thin dashed line the value
of this difference when not including the contribution
(\ref{ab3}).}\label{figAB}
\end{figure}

	We check in Fig.~\ref{figAB} the very good agreement with the Fourier
transform of the spectrum in vicinity of the length of the families drawn on
Fig.~\ref{AB_rect}a. In this Figure, the numerical $F(x)$ is computed using
Eq.~(\ref{f1bis}) with $\beta=5$, $k_{min}$ and $k_{max}$ being respectively
the first and the $1400^{th}$ level. In the numerical computation we took
$\alpha=1/2$ because in this case the diffractive effects on the level density
are at maximum (see Eq.~(\ref{ab3})). The shaded area hardly seen at the
bottom of the plot is the modulus of the difference between the numerical and
analytical $F(x)$. For obtaining the excellent agreement of Fig.~\ref{figAB}
we have taken into account classical isolated boundary orbits (of the type
already encountered for the family drawn in Fig.~\ref{frontiere3}) and simple
nearby diffractive periodic orbits which can be treated within the geometrical
theory of diffraction (the relevant formulae are given in \cite{sie99}). Not
taking into account the diffractive contribution (\ref{ab3}) would give a much
larger error which is represented by a thin dashed line.

\section{Conclusion}\label{conclusion}

 In this paper we have studied diffractive corrections to the semiclassical
trace formula for the level density of polygonal billiards. Special care has
been devoted to the treatment of diffractive periodic orbits lying on (or in
vicinity of) the optical boundary, i.e. on the verge of being allowed by
classical mechanics. In particular we derived a systematic expansion for the
corner-diffractive corrections to the $n^{th}$ iterate of a family of periodic
orbits.

	The method employed (based on approximation (\ref{e4})) allows to treat a
rich variety of different cases with great precision (Secs. \ref{section2} to
\ref{section7}). This method is easily extended to similar diffraction
problems. In particular, our approach to the diffractive correction of the
$n^{th}$ iterate of a family allows to treat the non-trivial problem of
multiple forward Aharonov-Bohm diffusion (Sec.~\ref{section8}).

	The main purpose of our study was to establish the basis of a trace formula
in pseudo-integrable systems, with contributions from diffractive orbits. It
seems that these diffractive corrections are responsible for particular forms
of spectral statistics observed in many such models \cite{bog99}. Further
investigations will elucidate this relationship.

\

\noindent {\large \bf Acknowledgments}

\bigskip \noindent It is a pleasure to thank M. Sieber for fruitful comments
on the manuscript and for bringing Ref.~\cite{dur88} to our attention.

\appendix
\section{}\label{unif}
\setcounter{equation}{0}

	In this Appendix we derive Kirchhoff-like formulae for the Green function in
the cases of corner and flux line diffraction (Eqs.~(\ref{e4},\ref{e5})). Here
we compare Eqs. (\ref{e4}) and (\ref{e5}) with the exact diffraction in the
free plane~: diffraction by an infinite wedge (in \ref{cd}) and by a flux line
in the plane (in \ref{fld}). Adding boundaries to the problem (e.g. putting
the flux line in a billiard) amounts -- through the method of images -- to add
other sources of diffraction. In this case we describe multiple diffraction by
using a natural generalization of Eqs. (\ref{e4},\ref{e5}) (see e.g.
Eq.~(\ref{f2})).

\subsection{Corner diffraction}\label{cd}

A uniform approximation for diffraction on a single corner has been first
given by Pauli \cite{pau38}. The subject has been studied in detail in the
late 1960s and in the 1970s. We state here the result of one possible method
of ``uniformization'' (detailed derivation and references can be found in
\cite{sie97})~:

\begin{equation}\label{a1}
G_{1d}(\vec r,\vrp,E)=\frac{1}{4}\, 
\frac{\ep^{i kL_d+i\pi/4}}{\sqrt{\pi k L_d}}
\sum_{\sigma , \eta =\pm 1} |a_{\sigma , \eta}| \; {\cal D}_{\sigma , \eta}
\; K\left[ |a_{\sigma , \eta}| \left(\frac{k \ell
\ell'}{L_d}\right)^{1/2}\right] \; , \end{equation}

\noindent where $\ell'$ and
$\ell$ are the lengths of the classical trajectories from $\vrp$ to $\vr0$
and from $\vr0$ to $\vec r$ ($L_d=\ell' + \ell$). 

$K$ is a modified Fresnel function which is defined by

\begin{equation}\label{a2}
K(z) = \frac{\ds 1}{\sqrt{\pi}} \, \exp\left\{-iz^2-i\pi/4\right\}
\int_{z}^{\infty} dt \, e^{i t^2} = 
\frac{\ep^{-i z^2}}{\ds 2} \;\mbox{erfc}\; (\ep^{-i\pi/4}\, z) \; ,
\end{equation}

and which has the following limiting properties : $K(0)=1/2$ and

\begin{equation}\label{a3}
K(z)\approx \frac{\ep^{i\pi/4}}{\ds 2 z \sqrt{\pi}}\; 
(1 - \frac{i}{2\,z^2} - \frac{3}{4\,z^4} \ldots ) 
\quad\mbox{when}\quad |z|\to + \infty
\quad\mbox{with}\quad -\frac{\pi}{4} < \;\mbox{arg}\; (z) < \frac{3\pi}{4}
\; .\end{equation}

	In (\ref{a1}), $a_{\sigma , \eta}$ is a kind of 
a measure of the angular
distance from the trajectory to the optical boundary. On the optical boundary
characterized by $\sigma$ and $\eta$, $a_{\sigma , \eta}=0$. Far from the
optical boundary its precise value is irrelevant since one can use
Keller's approximation (which corresponds to keeping only
the first term in expansion (\ref{a3})). In the transition region one has to
use a specific form of $a_{\sigma , \eta}$, which characterizes the type of
uniform approximation chosen. We take here (see \cite{sie97})

\begin{equation}\label{a4}
a_{\sigma , \eta} = \sqrt{2}\, 
\cos (\frac{\phi_\sigma}{\ds 2} - n_{\sigma ,\eta} \, \gamma ) 
\qquad\mbox{where}\qquad 
n_{\sigma , \eta} = \,\mbox{nint}\;
\left[\frac{\phi_\sigma+\eta\pi}{\ds 2\gamma}\right] \in \ZZ \; ,
\end{equation}

\noindent nint denoting the nearest integer and
$\phi_\sigma=\theta'-\sigma\theta$, where $\theta'$ ($\theta$) is the incoming
(outgoing) angle of the diffractive trajectory with the boundary.

\

	In the following of this Appendix, we will use the uniform
approximation (\ref{a1}) for justifying the approximation
(\ref{e4}) which is valid in vicinity of an optical boundary.
Let's consider that for one of the four couples of values
of $(\sigma,\eta)$ one is near the optical boundary. In the contribution of
the three other terms in (\ref{a1}), the modified Fresnel function 
can be evaluated by 
keeping only
the first term in expansion (\ref{a3}) and this gives the second
contribution in the r.h.s. of (\ref{e4}).

\begin{figure}[thb]
\begin{center}
\includegraphics*[width=12cm,bbllx=10pt, bblly=140pt, bburx=580pt,
bbury=500pt]{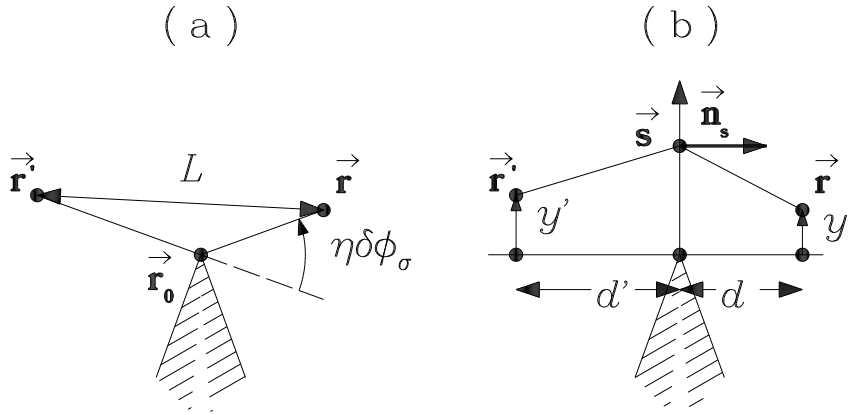}
\end{center}
\caption{\small Representation of a diffractive trajectory going from $\vrp$
to $\vec r$. Part (a) displays the notations of Eq.~(\ref{a5}) and part (b) of
Eq.~(\ref{a6}). In the figure, $\eta\,\delta\phi_\sigma >0$ (or equivalently
$y'/d'+y/d>0$) and the classical trajectory from $\vrp$ to $\ver$ is allowed.}
\label{fig_app}
\end{figure}

	For obtaining the first contribution of the r.h.s. of (\ref{e4}) one needs
to make explicit computations. The configuration we study has been represented
on Fig.~\ref{fig_app} for an orbit
near the optical boundary (the optical boundary for trajectories issued from
$\vrp$ is the dashed line of Fig.~\ref{fig_app}a). Note (from
Ref.~\cite{sie97}) that the classical orbit on the optical boundary has
properties depending on $\sigma$ and $\eta$. If $\sigma=1$ ($-1$) it has an
even (odd) number of bounces near the corner~; if $\eta=1$ ($-1$) it bounces
first on the line $\t=\gamma$ ($0$). If one writes $\phi_\sigma=\phi_{\sigma
0}+\delta\phi_\sigma$ (where $\phi_{\sigma 0}=2 n_{\sigma ,\eta} \,\gamma -
\eta\,\pi$ is the value of $\phi_\sigma$ on the optical boundary), by
examining the four different configurations, one can convince oneself
geometrically on Fig.~\ref{fig_app} that the oriented angle between
$(\vr0-\vrp)$ and $(\vec r-\vr0)$ is $\eta\,\delta\phi_\sigma$. If
$\eta\,\delta\phi_\sigma<0$, there is no classical orbit from $\vrp$ to $\vec
r$. If $\eta\,\delta\phi_\sigma>0$, the classical orbit is allowed and it has
a Maslov index $\nu$ such that $\exp\{i\nu\pi/2\}=\sigma$.
Since one is near the optical boundary, the angle $\delta\phi_\sigma$ is small
and in (\ref{a1}) one can make the approximations $a_{\sigma ,\eta} \approx
\eta\, \delta\phi_\sigma/\sqrt{2}$, ${\cal D}_{\sigma, \eta}\approx
2\sigma\eta/\delta\phi_\sigma$ (compare with the exact formulae (\ref{a4}) and
(\ref{e3})). One has also $\ell+\ell'-|a_{\sigma
,\eta}|^2\ell\ell'/L_d\approx L$, where $L$ is the length of the classical
path from $\vrp$ to $\vec r$. Altogether one obtains from (\ref{a1})~:

\begin{eqnarray}\label{a5}
G_{1d}(\vec r,\vrp,E)& \approx &  
\,\mbox{sgn}\, (\eta\,\delta\phi_\sigma)\;
\frac{\ds\ep^{i k L -i\nu\pi/2}}{2\pi\sqrt{\ds 2 k L}} \;
\int_{\sqrt{k(L_d-L)}\;}^{+\infty}\, \ep^{\ds i t^2} dt \nonumber\\
& & \nonumber\\
& + & G_{0}(\vec r,\vr0,E)\,
{\cal D}_{reg} (\t,\tp)\, G_{0}(\vr0,\vrp,E) \; ,
\end{eqnarray}

\noindent and

\begin{equation}\label{a5bis}
G_1(\vec r,\vrp,E)=G_{1d}(\vec r,\vrp,E) + \Theta (\eta\,\delta\phi_\sigma)\;
G_{0}(\vec r,\vrp,E)\; ,\end{equation}

\noindent where $\Theta$ is the Heaviside function.

	We will now show that this expression matches Eq. (\ref{e4}). For that
purpose we will explicitly evaluate Eq. (\ref{e4}) by choosing a particular
axis of coordinates, shown in Fig.~\ref{fig_app}(b). In that figure we have
chosen the locus of points $\vs$ such that the distance from $\vrp$ and $\vec
r$ to its perpendicular is small (this is consistent with the fact that the
trajectory is near the optical boundary). If one denotes by $d$ ($d'$) the
distance between $\vr0$ and the projection of $\vec r$ ($\vrp$) on the
perpendicular and by $y$ ($y'$) the algebraic distance from $\vec r$ ($\vrp$)
to the perpendicular (see Fig.~\ref{fig_app}(b)), one has~:

\begin{eqnarray}\label{a6}
-2 \int_0^{+\infty}\!\!\!ds\, G_{0}(\vs,\vrp,E) 
\vec{n}_s \cdot \vec{\nabla}_{\vs} G_{0}(\vec r,\vs,E)
&\approx& - \frac{\ep^{i k (d+d')-i\nu\pi/2}}{4\pi}
\int_0^{+\infty}\!\!\!\!\! ds \; 
\frac{\ep^{i \frac{k}{2}
\left\{\frac{(s-y')^2}{d'} +\frac{(s-y)^2}{d}\right\} }}
{\sqrt{\ds d \, d'}} \nonumber\\
& & \nonumber\\
&\approx& - \frac{\ep^{i k L-i\nu\pi/2}}{2\pi\sqrt{\ds 2 k L}}
\int_{-\,\mbox{\scriptsize sgn}\,
(\eta\,\delta\phi_\sigma)\,\sqrt{k(L_d-L)}\;}
^{+\infty}\, \ep^{\ds i t^2} dt
\; ,\end{eqnarray}

\noindent where one has made the change of variable
$t=\sqrt{k(d+d')/(2\,d\,d')} [s-(d y'+y d')/(d+d')]$ and one has used
the facts that $L_d-L\approx (d\,d'/2)(y'/d'+y/d)^2/(d+d')$ and
$\mbox{sgn}\,(y'/d'+y/d)=\,\mbox{sgn}\,(\eta\,\delta\phi_\sigma)$. This
expression inserted into (\ref{e4}) is equivalent to expression
(\ref{a5bis}) for the Green function.
Hence, starting from the uniform approximation (\ref{a1}), we have proven
the validity of Eq.~(\ref{e4}) near the optical boundary.

\subsection{Flux line diffraction}\label{fld}

	In the case of diffraction by a flux line, a uniform solution has been
worked out by Aharonov and Bohm \cite{aha59}. The corresponding expression for
the Green function, as given by Sieber \cite{sie99}, reads~:

\begin{equation}\label{a11}
G_1(\vec{r},\vrp,E) = G_0(\vec{r},\vrp,E) \, \ep^{i\alpha(\phi-\phi')}
+\sin(\alpha\pi)\,
\frac{\ep^{ik L_d+ i (\phi-\phi')/2+i\pi/4}}{\sqrt{2\pi k L_d}}
K \left[ \sqrt{\frac{2 k \ell \ell'}{L_d}} \, 
\cos\left(\frac{\phi-\phi'}{2}\right)\right]
\; ,\end{equation}

\noindent where $\phi$ and $\phi'$ are here the angular coordinates of a polar
system with origin on the flux line $\vr0$. They have to be chosen such that
$|\phi'-\phi|<\pi$. The other notations are identical to those of
Eq.~(\ref{a1}). In the configuration illustrated in Fig.~\ref{fig_gene} this
means that if $\delta\phi$ is the angle between $\vec{r}-\vr0$ and $\vr0-\vrp$
($\delta\phi$ is the analogous of $\eta\delta\phi_\sigma$ of
Fig.~\ref{fig_app}a) one has $\phi-\phi' = \delta\phi-\pi\,
\mbox{sgn}\,(\delta\phi)$. Note also that as in Ref.~\cite{sie99} we restrict
ourself to the case of non-singular behavior near the flux line (i.e.
vanishing wave-functions).

\

	We will now follow the same procedure as in \ref{cd} and show that
Eq.~(\ref{e5}) and (\ref{a11}) are equivalent in the limit of small
$\delta\phi$. In this limit the geometrical theory of diffraction fails, and
indeed it is well known that the Aharonov-Bohm scattering amplitude diverges
in the forward direction. The computations are very similar and for simplicity
we chose here the arbitrary locus of points $\vec{s}$ perpendicular to the
optical boundary which is the line going from $\vrp$ to $\vr0$, then $y'=0$
(the notations are defined in Fig.~\ref{fig_app}b). In the semiclassical
limit, Eq. (\ref{e5}) reads~:

\begin{equation}\label{a12}
G_1(\vec{r},\vrp,E) \approx - 
\frac{\ep^{\ds i k (d+d')+i \alpha\delta\phi }}{4\pi\sqrt{d d'}}
\int_{-\infty}^{+\infty}\!\!\!\! ds
\; \ep^{-i\,\mbox{\small sgn}\,(s) \pi} \;
\ep^{\frac{i k}{2}\left[\frac{s^2}{d'}+\frac{(s-y)^2}{d}\right]}
\; .
\end{equation}

	Using the fact that $L\approx d+d'+(1/2)y^2/(d+d')$ and
$L_d\approx d+d'+y^2/(2 d)$ this can be rewritten as~:

\begin{equation}\label{a13}
G_1(\vec{r},\vrp,E) \approx - 
\frac{\ep^{\ds i (k L +\alpha\delta\phi)}}{\pi\sqrt{8 k (d+d')}}
\left\{
\ep^{-i\alpha\pi}
\int_{-\epsilon\sqrt{k\Delta}}^{+\infty}\!\!dt \; \ep^{i t^2}
+ \ep^{i\alpha\pi}
\int^{-\epsilon\sqrt{k\Delta}}_{-\infty}\!\!\!\!\! dt\;  \ep^{i t^2}
\right\} \; ,
\end{equation}

\noindent where $\Delta=L_d-L$ and $\epsilon=\,\mbox{sgn}\,(\delta\phi)
=\,\mbox{sgn}\,(y)$. Simple manipulations show that

\begin{equation}\label{a14}
G_1(\vec{r},\vrp,E) \approx G_0(\vec{r},\vrp,E) \; 
\ep^{i\alpha(\delta\phi-\epsilon\pi)}
+\frac{\sin(\alpha\pi)}{\sqrt{2\pi k L_d}}
\ep^{i(k L_d+\pi/4-\epsilon\pi/2)} K\left(\sqrt{k \Delta}\right) \; ,
\end{equation}

\noindent and this expression matches (\ref{a11}) when $|\delta\phi|<<\pi$.

\section{}\label{integrales}
\setcounter{equation}{0}

	In this Appendix we give some useful formulae that correspond to
transverse integration of the different types of Green functions appearing
in the main text.

\

	Let's first define $F(x,y,s_1,\ldots s_n)$ by
	
\begin{equation}\label{b2}
F(x,y,s_1,\ldots ,s_n) =
\exp\left\{\frac{i k}{2}\left[
\frac{(s_1-y)^2}{x} + \frac{(s_2-s_1)^2}{\ell_2} + ...
+\frac{(s_n-s_{n-1})^2}{\ell_n}+\frac{(y-s_n)^2}{\ell_1-x}
\right]\right\}
\; ,\end{equation}

\noindent and $D_n(x,\ell_1,\ldots \ell_n)$ by

\begin{equation}\label{b3}
D_n(x,\ell_1,\ldots , \ell_n)
=\frac{\ds \ep^{i k L -3 (n+1) i \pi/4}}
{(\ds 8\pi k)^{\frac{n+1}{2}}
\sqrt{\ds x(\ell_1-x)\ell_2\ldots \ell_n}} \; ,
\end{equation}

\noindent where $L=\ell_1+\ldots + \ell_n$ and $0\le x\le \ell_1$.

\

	$\bullet$ In Section \ref{front}, for treating the first order diffractive
correction to the $n^{th}$ repetition of a family, one needs to compute the
following integral (cf. e.g. Eq. (\ref{f3}) which after transverse integration
yields the expression of $I_2(\ell,\ell)$) ~:

\begin{eqnarray}\label{b4}
I_n(\ell_1,\ldots ,\ell_n)
=(2 i k)^n\, D_n(x,\ell_1,\ldots \ell_n)
\int_{-\infty}^{+\infty}\!\!\!\!\!\!\!\!\! & dy &
\!\!\!\!\!\left[
\int_0^{+\infty}\!\!\!\!\! ...\int_0^{+\infty}\!\!\!\!\! ds_1\ldots ds_n\;
F(x,y,s_1,\ldots s_n)
\right.\nonumber\\
& & \nonumber\\
& - & \!\!\!\!\! \left. \Theta(y)
\int_{-\infty}^{+\infty}\!\!\!\!\! 
...\int_{-\infty}^{+\infty}\!\!\!\!\! ds_1\ldots ds_n\;
F(x,y,s_1,\ldots s_n)
\right] \; .
\end{eqnarray}

	It is easy to see that the $x$ dependence disappears
in expression (\ref{b4}) after integration over $y$. This is the reason why
we did not include $x$ in the list of arguments of $I_n$.

\

	We obtain
	
\begin{equation}\label{b5}
I_2(\ell_1,\ell_2)=
\frac{\ep^{i k L+i\pi/2}}{4\pi k L}\, \sqrt{\ell_1\ell_2}\; ,
\end{equation}

\begin{equation}\label{b6}
I_3(\ell_1,\ell_2,\ell_3)=
\frac{\ep^{i k L+i\pi/2}}{8\pi k L}\, \left\{
\sqrt{\ell_1(\ell_2+\ell_3)} +
\sqrt{\ell_2(\ell_3+\ell_1)} +
\sqrt{\ell_3(\ell_1+\ell_2)} \right\}
\; ,
\end{equation}

\begin{eqnarray}\label{b7}
I_4(\ell_1,\ell_2,\ell_3,\ell_4) & = &
\frac{\ep^{i k L+i\pi/2}}{16\pi k L}\, \left\{
\sqrt{\ell_1(\ell_2+\ell_3+\ell_4)} +
\sqrt{\ell_2(\ell_3+\ell_4+\ell_1)} +
\sqrt{\ell_3(\ell_4+\ell_1+\ell_2)}\right.\nonumber\\
& & \nonumber\\
& + &
\sqrt{\ell_4(\ell_1+\ell_2+\ell_3)} +
\sqrt{(\ell_1+\ell_2)(\ell_3+\ell_4)} +
\sqrt{(\ell_1+\ell_4)(\ell_2+\ell_3)} \nonumber\\
& & \nonumber\\
& + &
\frac{2}{\pi}\sqrt{\ell_1(\ell_2+\ell_3+\ell_4)}
\,\mbox{arctg}\,\sqrt{\frac{\ds
\ell_2\ell_4}{\ds\ell_3(\ell_2+\ell_3+\ell_4)}}\nonumber\\
& & \nonumber\\
& + &
\frac{2}{\pi}\sqrt{\ell_2(\ell_3+\ell_4+\ell_1)}
\,\mbox{arctg}\,\sqrt{\frac{\ds
\ell_3\ell_1}{\ds\ell_4(\ell_3+\ell_4+\ell_1)}}\nonumber\\
& & \nonumber\\
&+&
\frac{2}{\pi}\sqrt{\ell_3(\ell_4+\ell_1+\ell_2)}
\,\mbox{arctg}\,\sqrt{\frac{\ds
\ell_4\ell_2}{\ds\ell_1(\ell_4+\ell_1+\ell_2)}}\nonumber\\
& & \nonumber\\
& + &
\left.
\frac{2}{\pi}\sqrt{\ell_4(\ell_1+\ell_2+\ell_3)}
\,\mbox{arctg}\,\sqrt{\frac{\ds
\ell_1\ell_3}{\ds\ell_2(\ell_1+\ell_2+\ell_3)}}
\right\}\; ,\end{eqnarray}

\noindent and

\begin{eqnarray}\label{b8}
2\,I_5(\ell_1,\ell_2,\ell_3,\ell_4,\ell_5) & = &
I_4(\ell_1+\ell_5,\ell_2,\ell_3,\ell_4)
+I_4(\ell_1+\ell_2,\ell_3,\ell_4,\ell_5)
+I_4(\ell_1,\ell_2+\ell_3,\ell_4,\ell_5)\nonumber\\
& & \nonumber\\
& + &
I_4(\ell_1,\ell_2,\ell_3+\ell_4,\ell_5)
+I_4(\ell_1,\ell_2,\ell_3,\ell_4+\ell_5)
-I_3(\ell_1+\ell_2+\ell_5,\ell_3,\ell_4)\nonumber\\
& & \nonumber\\
& - &
I_3(\ell_1,\ell_2+\ell_3+\ell_4,\ell_5)
-I_3(\ell_1,\ell_2+\ell_3,\ell_4+\ell_5)
-I_3(\ell_1,\ell_2,\ell_3+\ell_4+\ell_5)\nonumber\\
& & \nonumber\\
& + &
I_2(\ell_1,\ell_2+\ell_3+\ell_4+\ell_5)
\; .\end{eqnarray}

	Although this is not apparent in the above expression, explicit computation
shows that formula (\ref{b8}) is -- as Eqs.~(\ref{b5},\ref{b6},\ref{b7}) --
invariant under cyclic permutation of the indices.

	Expressions (\ref{b5}) to (\ref{b8}) greatly simplify when all the $\ell$'s
are equal. One obtains~:

\begin{equation}\label{b8bis}
I_n(\ell, \ldots \ell)=\frac{C_n}{k}\, \ep^{\ds i k L+i\pi/2}
\; ,\end{equation}

\noindent with $C_1=0$, $C_2=1/(8\,\pi)$, $C_3=1/(4\pi\sqrt{2})$,
$C_4=(1+4/\sqrt{3})/(16\,\pi)$, $C_5=(1+\sqrt{6}/3)/(8\,\pi)$. A general
formula for $C_n$ is given in \ref{valeurcn}. From (\ref{b8bis}) the
contribution (\ref{f7}) to the level density follows directly.

\

	$\bullet$ The next order correction to (\ref{b4}) requires the
	computation of the following integral~:

\begin{eqnarray}\label{b9}
J_n(\ell_1,\ldots ,\ell_n)
& = & (2 i k)^{n-1}\, D_n
\int_{-\infty}^{+\infty}\!\!\!\!\! dy 
\left[ {\cal D}^{(1)}_{reg} 
\int_0^{+\infty}\!\!\!\!\! ...\int_0^{+\infty}\!\!\!\!\! ds_2\ldots ds_n\;
F(x,y,0,s_2,\ldots , s_n) \right. \nonumber\\
& & \nonumber\\
&+& {\cal D}^{(2)}_{reg}
\int_0^{+\infty}\!\!\!\!\! ...\int_0^{+\infty}\!\!\!\!\! ds_1ds_3\ldots ds_n\;
F(x,y,s_1,0,s_3\ldots , s_n) + \ldots \nonumber \\
&+& \left. {\cal D}^{(n)}_{reg}
\int_0^{+\infty}\!\!\!\!\! ...\int_0^{+\infty}\!\!\!\!\! ds_1\ldots ds_{n-1}\;
F(x,y,s_1,\ldots , s_{n-1},0) \right]
 \; .
\end{eqnarray}

	This expression corresponds to a sum of $n$ trajectories~; the $j^{st}$
trajectory having one Keller bounce on apex $j$ (with a regular part of the
diffraction coefficient noted ${\cal D}^{(j)}_{reg}$, $j=1\ldots n$) and
Kirchhoff contributions from the other apexes (for instance Eq.~(\ref{f5})
corresponds after transverse integration to $J_2(\ell,\ell)$).

\

	We obtain

\begin{equation}\label{b10}
J_2(\ell_1,\ell_2) = ({\cal D}^{(1)}_{reg}+{\cal D}^{(2)}_{reg})\,
\frac{\ep^{i k L +3i\pi/4}}{4k\sqrt{\ds 8\pi kL}}
\end{equation}

and

\begin{eqnarray}\label{b11}
J_3(\ell_1,\ell_2,\ell_3) & = &
\frac{\ep^{i k L +3i\pi/4}}{8 k\sqrt{\ds 8\pi kL}} \left[
{\cal D}^{(1)}_{reg} \left( 1+\frac{2}{\pi}
\,\mbox{arctg}\,
\sqrt{\frac{\ds\ell_1\ell_2}{\ds\ell_3 L}} \, \right) \right.\nonumber \\
& & \nonumber\\
& + & 
\left. {\cal D}^{(2)}_{reg} \left( 1+\frac{2}{\pi}
\,\mbox{arctg}\,
\sqrt{\frac{\ds\ell_2\ell_3}{\ds\ell_1 L}} \, \right) +
{\cal D}^{(3)}_{reg} \left( 1+\frac{2}{\pi}
\,\mbox{arctg}\,
\sqrt{\frac{\ds\ell_3\ell_1}{\ds\ell_2 L}} \, \right) \right]
\end{eqnarray}

	The expression of $J_n$ when all the $\ell$'s are equal is~:

\begin{equation}\label{b12}
J_n(\ell,... , \ell) = - \frac{\ep^{i k L - i n\pi/4}}
{4\pi^{n/2}k\sqrt{2k\ell}} \, n\,{\cal D}_{reg}\,  \tilde{J}_{n-1}\; ,
\end{equation}

\noindent where

\begin{equation}\label{b12bis}
\tilde{J}_n =
\int_0^{+\infty}\!\!\!\!\! ...\int_0^{+\infty}\!\!\!\!\!dx_1...dx_n\;
\ep^{i\left[x_1^2+(x_2-x_1)^2+...+(x_n-x_{n-1})^2+x_n^2\right]}
\; .
\end{equation}

	It is shown in \ref{eugene} that $\tilde{J}_n=(\ep^{i\pi/4}
\sqrt{\pi})^n\,(n+1)^{-3/2}$. From this result and Eq. (\ref{b12}), formula
(\ref{f6}) follows immediately.

\

	$\bullet$ In section \ref{deuxfront}, in order to compute the first order
correction to the contribution of a family whose boundary partly
coincides with the frontier of the billiard, one needs to compute the
following integral~:

\begin{equation}\label{b13}
M_2(\ell_1,\ell_2) = I_2(\ell_1,\ell_2)
- (2 i k)^2 D_2\int_{-\infty}^{+\infty}\!\!\!\!\!dy
\int_0^{+\infty}\!\!\!\!\!ds_1\int_0^{+\infty}\!\!\!\!\!ds_2\, 
\exp\left\{\frac{i k}{2}\left[
\frac{(s_1-y)^2}{x}+\frac{(s_2+s_1)^2}{\ell_2}+\frac{(y-s_2)^2}{\ell_1-x}
\right]\right\}
 \; .
\end{equation}

	This equation corresponds to the transverse integration of (\ref{df1}) after
the removing of the contribution of the the direct path (i.e. of $G_0$). The
last integral in the r.h.s. of (\ref{b13}) corresponds to the orbit going from
$\ver$ to $\ver$ and bouncing on the boundary of the family which is also a
frontier of the billiard. The term $I_2(\ell_1,\ell_2)$ corresponds to the
direct diffractive trajectory. One obtains

\begin{equation}\label{b14}
M_2(\ell_1,\ell_2)= \frac{\ep^{i k L+i\pi/2}}{4\pi k L}
\left( \sqrt{\ell_1 \ell_2} +
L\,\mbox{arctg}\, \sqrt{\frac{\ell_2}{\ell_1}}\,\right) \; .
\end{equation}

	Here we want to develop a point stated in the main text~: if $\ver$ lies
near the part of the optical boundary that coincides with the frontier of the
billiard, the Green function has four contributions which, after integration
over $y$, give the same contribution as (\ref{b14}).

  If one defines

\begin{eqnarray}\label{b150}
h^1(x,y,s_1,s_2) & = &
\exp\left\{\frac{i k}{2}\left[
\frac{(s_2-y)^2}{x}+\frac{(s_1-s_2)^2}{\ell_1}+\frac{(y-s_1)^2}{\ell_2-x}
\right]\right\}
\; ,\nonumber\\
& & \nonumber\\
h^2(x,y,s_1,s_2) & = &
\exp\left\{\frac{i k}{2}\left[
\frac{(s_2+y)^2}{x}+\frac{(s_1-s_2)^2}{\ell_1}+\frac{(y-s_1)^2}{\ell_2-x}
\right]\right\}
\; ,\nonumber\\
& & \nonumber\\
h^3(x,y,s_1,s_2) & = &
\exp\left\{\frac{i k}{2}\left[
\frac{(s_2+y)^2}{x}+\frac{(s_1-s_2)^2}{\ell_1}+\frac{(y+s_1)^2}{\ell_2-x}
\right]\right\}
 \; ,\nonumber\\
& & \nonumber\\
h^4(x,y,s_1,s_2) & = &
\exp\left\{\frac{i k}{2}\left[
\frac{(s_2-y)^2}{x}+\frac{(s_1-s_2)^2}{\ell_1}+\frac{(y+s_1)^2}{\ell_2-x}
\right]\right\}
\; ,
\end{eqnarray}

\noindent then, for a point $\ver$ near the boundary of the orbit
which coincides with the frontier of the billiard,
the four contributions to the Green function integrated
transversely to the direction of the orbit read~:

\begin{equation}\label{b15}
M_2^1(x)=
(2 i k)^2 D_2(x,\ell_2,\ell_1)
\int_0^{+\infty}\!\!\!\!\!dy\left[
\int_0^{+\infty}\!\!\!\!\!ds_1\int_0^{+\infty}\!\!\!\!\!ds_2\,
 h^1(x,y,s_1,s_2)
- \int_{-\infty}^{+\infty}\!\!\!\!\!ds_1
\int_{-\infty}^{+\infty}\!\!\!\!\!ds_2\,
 h^1(x,y,s_1,s_2)\right] \; ,
\end{equation}

\noindent and

\begin{equation}\label{b16}
M_2^j(x)= (-1)^{j+1}\, (2 i k)^2 D_2(x,\ell_2,\ell_1)
\int_0^{+\infty}\!\!\!\!\!dy
\int_0^{+\infty}\!\!\!\!\!ds_1\int_0^{+\infty}\!\!\!\!\!ds_2\,
 h^j(x,y,s_1,s_2) \; ,
\quad\mbox{for}\quad j=2,3,4 \; .
\end{equation}

\begin{figure}[thb]
\begin{center}
\includegraphics*[width=10cm,bbllx=50pt, bblly=80pt, bburx=575pt,
bbury=520pt]{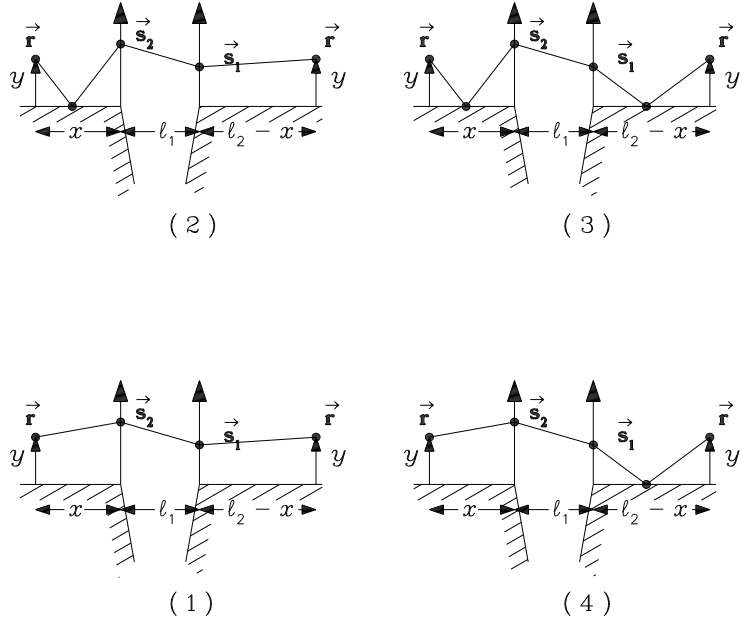}
\end{center}
\caption{\small Schematic representation of the path encompassed in the
contribution $M_2^j$ (Eqs. (\ref{b15},\ref{b16})) to the transverse
integration of the Green function ($j=1...4$). The plot labelled $(j)$
corresponds to $M_2^j(x)$. One has here four different contributions because
the initial point $\vec{r}$ lies along the part of the boundary of the family
that coincides with the frontier of the billiard. The simpler case that
$\vec{r}$ is not in vicinity of the frontier of the billiard is represented in
Fig. \ref{deux_front2}.}
\label{deux_front3}
\end{figure}

	Note that the transverse integration (over the variable $y$) is only
possible here from $0$ to $+\infty$ because one is near the frontier of the
billiard (see Fig. \ref{deux_front3}). The four contributions
(\ref{b15},\ref{b16}) correspond to different paths going from $\ver$ to
$\ver$~: $M_2^1$ corresponds to a path going from $\ver$ to $\vec{s_2}$, to
$\vec{s}_1$ and back to $\ver$ (for this part one has to withdraw the
semi-classical Green function)~; $M_2^2$ corresponds to the path going from
$\ver$ to $\vec{s}_2$ with a reflection on the boundary of the orbit which
coincides with the frontier of the billiard, then going from $\vec{s}_2$ to
$\vec{s}_1$ and back to $\ver$ etc ... This is illustrated in Figure
\ref{deux_front3}.

\
 
	The $M_2^j$'s separately depend on $x$, but one obtains~:

\begin{equation}\label{b17}
M_2^1(x)+M_2^3(x) =
\frac{\ep^{i k L+i\pi/2}}{4\pi k L}\,\sqrt{\ell_1\ell_2}
\quad\mbox{and}\quad
M_2^2(x)+M_2^4(x) =
\frac{\ep^{i k L+i\pi/2}}{4\pi k}
\,\mbox{arctg}\,\sqrt{\frac{\ds\ell_2}{\ds\ell_1}}\; .
\end{equation}

	Hence, when $\ver$ lies near the frontier of the billiard, one obtains
$M_2^1(x)+M_2^2(x)+M_2^3(x)+M_2^4(x)=M_2$, where $M_2$ is given by (\ref{b14})
and corresponds to transverse integration when $\ver$ is not in vicinity of
the frontier of the billiard. Thus the result of the transverse integration of
the Green function does not depend on the position $x$ of the point $\ver$
along the orbit. This directly leads to formula (\ref{df2}).

\

	$\bullet$ In Section \ref{updown}, for evaluating the contribution of
a diffractive orbit jumping from one boundary of a family to another one needs
to compute the following integral (which is the transverse integration of
Eq.~(\ref{ud1}))~:

\begin{equation}\label{b18}
N_2(\ell_1,\ell_2) = (2 i k)^2 D_2(x,\ell_1,\ell_2)
\int^{+\infty}_{-\infty}\!\!\!\!\!dy  
\int_{-\infty}^\zeta\!\!\!\!\!ds_1
\int^{+\infty}_0\!\!\!\!\!ds_2\,
F(x,y,s_1,s_2)\; .\end{equation}

	The $y$ integration is trivial and (as now usual) removes the $x$
dependence. One obtains~:

\begin{eqnarray}\label{b19}
N_2(\ell_1,\ell_2)&=&- \frac{\ep^{i k L}}{4\pi\sqrt{\ds\ell_1\ell_2}}
\int^{+\infty}_0\!\!\!\!\!du_1
\int^{+\infty}_0\!\!\!\!\!du_2\,
\ep^{i \frac{k L}{2\ell_1\ell_2}(u_1+u_2-\zeta)^2}
\nonumber\\
& & \nonumber \\
&  = &        
\frac{\sqrt{\ds\ell_1\ell_2}}{4 i\pi k L} \ep^{i k L_d}
-\frac{\zeta\,\ep^{i k L}}{\sqrt{8\pi^2 k L}}
\int^{+\infty}_{-\sqrt{k\Delta}}\! du\, \ep^{iu^2}\nonumber\\
& & \nonumber\\
& =&
\frac{\sqrt{\ds\ell_1\ell_2}}{4 i\pi k L} \ep^{i k L_d}
-\frac{\zeta\,\ep^{i k L+i\pi/4}}{\sqrt{8\pi k L}}
\left[ 1 -\ep^{i k \Delta} K(\sqrt{k\Delta}\,)\right]
\; ,\end{eqnarray}

\noindent where $L_d$ is the length of the diffractive orbit close to the
family of length $L$ and $\Delta=L_d-L$ ($\Delta\simeq
\zeta^2\,L/(2\ell_1\ell_2)$). In the above expression, the last term simply
re-expresses the previous one using the modified Fresnel integral defined in
Eq.~(\ref{a2}). The longitudinal integration (which simply amounts to a
multiplication by $L$) of (\ref{b19}) yields the contribution to $\rho(E)$ of
the configuration studied in Section~\ref{updown}. More precisely, this
contribution reads $-(L/\pi)\, \mbox{Im}\, N_2(\ell_1,\ell_2)$. This results
in Eq.~(\ref{ud2}).

\section{}\label{eugene}
\setcounter{equation}{0}

	The purpose of this Appendix is the explicit computation of the integral
(\ref{b12bis})~:

\begin{equation}\label{d1}
\tilde{J}_n =
\int_0^{+\infty}\!\!\!\!\!...\int_0^{+\infty}\!\!\!\!\!
dx_1...dx_n\; \ep^{\ds i\, \Phi_n(\underline{x})}
\; ,
\end{equation}

\noindent where $\Phi_n(\underline{x})$ is the following quadratic form

\begin{equation}\label{d2}
\Phi_n(\underline{x})=x_1^2+(x_1-x_2)^2 + ... +(x_{n-1}-x_n)^2 + x_n^2 
\; .\end{equation}

	Note that in all this Appendix we denote $n$--dimensional vectors
$\underline{x} = (x_1,...,x_n)$. A key point in the evaluation
of integral (\ref{d1}) is the existence of a group generated by a set
of transformations
$\left\{T_j\right\}_{1\le j\le n}$ which leaves the quadratic form invariant :

\begin{equation}\label{d3}
T_j (\underline{x}) =\underline{x}' \quad\mbox{with}\quad
\left\{
\begin{array} {ccl}
x'_j & = & -x_j + x_{j+1} + x_{j-1} \; , \\
x'_k & = & x_k\;  , \qquad k\ne j \; ,\\
\end{array}
\right.
\end{equation}

\noindent where $j=1...n$ and we have adopted the convention
$x_{-1}=x_{n+1}=0$.

	These transformations are inversions ($T_j^2=1$) and they generate a finite
group (of the $A_n$ type, see e.g. \cite{cox65}). We give below a method of
calculation of (\ref{d1}) which does not require knowledge of the theory
of finite groups.

\

	$\bullet$ The quadratic form (\ref{d2}) can be naturally rewritten in the
form~:

\begin{equation}\label{d4}
\Phi_n(\underline{x})=2\, \sum_{i=1}^n M_{ij}\, x_i \, x_j
\end{equation}

\noindent where the $n\times n$ matrix $\underline{M}$ is the following~:

\begin{equation}\label{d5}
\underline{M} =\left(
\begin{array}{cccccccc}
1    & -1/2 & 0    & 0    & ... & 0    & 0    & 0    \\
-1/2 &   1  & -1/2 & 0    & ... & 0    & 0    & 0    \\
0    & -1/2 & 1    & -1/2 & ... & 0    & 0    & 0    \\
     &      &      &      & ... &      &      &      \\
0    & 0    & 0    & 0    & ... & -1/2 & 1    & -1/2 \\
0    &   0  & 0    & 0    & ... & 0    & -1/2 & 1    \\
\end{array}\right)\; .\end{equation}

\

	Due to the simple tri-diagonal structure of $\underline{M}$, it easy to
perform a Gaussian decomposition of the quadratic form $\Phi_n$ recursively.
This results in

\begin{equation}\label{d7}
\frac{1}{2}\Phi_n(\underline{x})=
\sum_{k=1}^{n-1} \lambda_k \, (x_k-\frac{x_{k+1}}{2\, \lambda_k})^2
+\lambda_n \, x_n^2
\; .\end{equation}

\noindent where

\begin{equation}\label{d8} 
\lambda_{k}=\frac{k+1}{2 k} \qquad k=1...n
\; .\end{equation}

	From (\ref{d7},\ref{d8}) it is clear that the determinant of $\underline{M}$
is

\begin{equation}\label{d6}
\mbox{det}\; \underline{M}
 = \lambda_1...\lambda_n = \frac{n+1}{\ds 2^n}
\; .\end{equation}

	Let us now introduce $n$ vectors $\underline{V}^j$ such that

\begin{equation}\label{d10}
M_{ij}=\underline{V}^i . \underline{V}^j \; .\end{equation}

	From (\ref{d5}) it is clear that (if these vectors exist) they are of unit
length and that the angle between different vectors equals either $\pi/2$
or $2\pi/3$ (the cosines being either 0 or $-1/2$). One possible
solution of Eqs. (\ref{d10}) can be written in the following form~:

\begin{eqnarray}\label{d11}
\underline{V}^1 & = & (\sqrt{\lambda_1},0, ... ,0) \; ,\nonumber \\
\underline{V}^2 & = & 
(\frac{-1}{2\sqrt{\lambda_1}},\sqrt{\lambda_2},0,...,0) \; ,\nonumber \\
   & ... & \nonumber\\
\underline{V}^k & = & 
(0,...,0,\frac{-1}{2\sqrt{\lambda_{k-1}}},\sqrt{\lambda_k},0,...,0)\; ,
\nonumber\\
   & ... & \nonumber\\
\underline{V}^n & = & 
(0,...,0,\frac{-1}{2\sqrt{\lambda_{n-1}}},\sqrt{\lambda_n})\; ,\nonumber\\ 
\end{eqnarray}

\noindent where the $\lambda_k$'s are defined in (\ref{d7},\ref{d8}). From
geometrical considerations it is clear that all other solutions can be
obtained from Eq. (\ref{d11}) by applying overall rotations (and inversions).

	Using the vectors $\underline{V}^j$ it is possible to define new coordinates
$(y_j)$ from the relation

\begin{equation}\label{d12}
\underline{y}=\sum_{j=1}^n x_j \, \underline{V}^j  \; .
\end{equation}

	Note that 

\begin{equation}\label{d12bis}
\underline{y}.\underline{y}=
\sum_{i,j=1}^n x_i\,x_j\, \underline{V}^i.\underline{V}^j =
\Phi_n(\underline{x})/2 \; .\end{equation}

	 The Jacobian of the transformation (\ref{d12}) is

\begin{equation}\label{d13}
{\cal J}=\,\mbox{det}\, \frac{\partial y_k}{\partial x_j} =
\,\mbox{det}\, V^j_k \; .\end{equation}

	It can be computed from Eqs. (\ref{d11}) or by taking the square of
expression (\ref{d13}). This yields ${\cal J}^2=\,\mbox{det}\,\underline{M}$.
Therefore

\begin{equation}\label{d14}
\int_0^{+\infty}\!\!\!\!\! ...\int_0^{+\infty}\!\!\!\!\!dx_1...dx_n\;
\ep^{\ds i\, \Phi_n(\underline{x})}
=\frac{1}{|\,\mbox{det}\,\underline{M}|^{1/2}} \;
\int\! ...\int_{\Omega}\!\! dy_1...dy_n\;
\ep^{\ds \, 2\, i\,(y_1^2+...+y_n^2)} \; ,
\end{equation}

\noindent where the integration is taken over the interior $\Omega$ of the
hyper-spherical simplex defined by the $n$ vectors $\underline{V}^j$, 

\begin{equation}\label{d15}
\Omega = \left\{ \underline{y} = \sum_{j=1}^n x_j \, \underline{V}^j
\quad\mbox{and}\quad  x_j\ge 0 \; ; \; j=1,...,n \right\} \; .
\end{equation}

	$\bullet$ The next and final step of the computation of (\ref{d1}) is to
show that the integration domain $\Omega$ in (\ref{d14}) is a relatively
simple sub-part of the whole $n$-dimensional space. To understand
geometrically the structure of this region it is convenient to add a new
vector $\underline{V}^{n+1}$ to the list (\ref{d11}) such that

\begin{equation}\label{d16}
\underline{V}^1+ ...+ \underline{V}^n+\underline{V}^{n+1} =0
\; .\end{equation}

	Angles formed by $\underline{V}^{n+1}$ with the other $\underline{V}^j$'s
are straightforwardly obtained from the expression (\ref{d5})~:
$\underline{V}^{n+1}.\underline{V}^1=
\underline{V}^{n+1}.\underline{V}^{n}=-1/2$,
$\underline{V}^{n+1}.\underline{V}^{j}=0$ for $j=2,...,n-1$ and
$\underline{V}^{n+1}.\underline{V}^{n+1}=\sum_{i,j}
\underline{V}^i.\underline{V}^j=1$.

	From the $(n+1)$ vectors $\underline{V}^{1}$, ...,$\underline{V}^{n+1}$ one
can define $(n+1)$ regions $\Omega_j$~:

\begin{equation}\label{d17}
\Omega_j = \left\{ \underline{y} = \sum_{j=1}^n x_j \, \underline{W}^j
\quad\mbox{and} \quad x_j\ge 0 \; ; \; j=1,...,n \right\} \; .
\end{equation}

\noindent where the $n$ vectors $\underline{W}^j$ include all $(n+1)$ vectors
$\underline{V}^k$ but the vector $\underline{V}^j$. In these notations the
region $\Omega$ in (\ref{d15}) coincides with $\Omega_{n+1}$. 

	It is almost evident that these $(n+1)$ regions cover the whole
$\underline{y}$ space without common intersection points. A formal proof of
this statement can be the following~:

	An arbitrary point $\underline{y}$ of the $n$-dimensional space has a unique
decomposition on the non-orthogonal basis of the $\underline{V}^j$'s
($j=1,...n)$ as given by (\ref{d12}). If all $x_j\ge 0$ then
$\underline{y}\in\Omega_{n+1}$, otherwise the set of coordinates $x_j$ is
divided into 2 sets ${x'_{\alpha}}$ and ${x''_{\beta}}$ of positive
($x'_{\alpha}\ge 0$) and negative ($x''_{\beta} < 0$) coordinates. Denoting by
$z_\beta=-x''_\beta$ and by $z_\gamma$ the maximum of the $z_\beta$'s we
get~:

\begin{equation}\label{d18}
\underline{y}=\sum_{\alpha}x_\alpha\,\underline{V}^\alpha
- \sum_{\beta\ne\gamma}z_\beta\,\underline{V}^\beta
- z_\gamma\,\underline{V^\gamma} \; .
\end{equation}

	Using the definition (\ref{d16}) one can express $\underline{V}^\gamma$ as a
function of the other $\underline{V}$'s and rewrite the above expression as

\begin{equation}\label{d19}
\underline{y}=\sum_{\alpha}(x_\alpha+z_\gamma)\,\underline{V}^\alpha
+\sum_{\beta\ne\gamma}(z_\gamma-z_\beta)\,\underline{V}^\beta
+z_\gamma \,\underline{V}^{n+1} \; .
\end{equation}

	As $z_\gamma\ge z_\beta$ all coefficients in this sum are non-negative and
the point $\underline{y}$ hence belongs to $\Omega_\gamma$. It is thus clear
that regions $\Omega_j$'s have no common points except on boundaries where
some $x_i=0$. It is also clear that the union of all the $\Omega_j$'s
($j=1...n+1$) covers all space since, given an arbitrary point
$\underline{y}$, one can assert unambiguously from the above procedure at
which of the $\Omega_j$'s it belongs.

	$\bullet$ Each region $\Omega_j$ is defined by $n$ vectors $\underline{W}^j$
obtained from the $(n+1)$ vectors $\underline{V}^k$'s by ignoring
$\underline{V}^j$. The convenient rearrangement of vectors $\underline{W}^j$'s
is the following~:

\begin{equation}\label{d20}
\left(\underline{W}^j\right)_{1\ge j\ge n}
=\left( \underline{V}^{j+1},
\underline{V}^{j+2},...,\underline{V}^n,\underline{V}^{n+1},
\underline{V}^1,\underline{V}^2, ...,\underline{V}^{j-1} \right)
\; .\end{equation}

	Let us now construct the matrix of mutual projections
$N_{ij}=\underline{W}^i.\underline{W}^j$. It is easy to check that this matrix
coincides with the matrix $\underline{M}$ defined in Eq. (\ref{d5}). Therefore
the vectors $\underline{W}^j$ $(j=1...n)$ will have the same mutual positions
as our initial vectors $\underline{V}^j$'s. As we noted above, it means that
region $\Omega_j$ for all $j$ $(j=1...n+1)$ can be obtained from the initial
region $\Omega\, (=\Omega_{n+1})$ by overall $n$-dimensional rotations (and
possibly by inversions). But the integrand in the r.h.s. of Eq. (\ref{d14}) is
invariant under such transformations, therefore its integration over any of
the $\Omega_j$'s is the same and~:

\begin{equation}\label{d21}
\int\! ...\int_{\Omega}\!\! dy_1...dy_n\;
\ep^{\, 2\, i\,(y_1^2+...+y_n^2)}
= \frac{1}{n+1}\,
\int_{-\infty}^{+\infty}\!\!\!\!\! ...
\int_{-\infty}^{+\infty}\!\!\!\!\!dy_1...dy_n\;
\ep^{\, 2\, i\,(y_1^2+...+y_n^2)} =
\frac{\left(\ep^{i\pi/4}\sqrt{\pi/2}\right)^n}{n+1}
\end{equation}

	Combining this result with Eqs. (\ref{d14}) and (\ref{d6}) one obtains

\begin{equation}\label{d22}
\tilde{J}_n =
\int_0^{+\infty}\!\!\!\!\! ...\int_0^{+\infty}\!\!\!\!\!dx_1...dx_n\;
\ep^{\ds i\, \Phi_n(\underline{x})}
=\frac{(\ep^{i\pi/4}\sqrt{\pi})^n}{(n+1)^{3/2}}
\; .\end{equation}

\noindent which is the result needed for the explicit computation of
expression (\ref{b12}). From the approach exposed in this Appendix, one can
state the more general result~:

\begin{equation}\label{d23}
\int_0^{+\infty}\!\!\!\!\! ...\int_0^{+\infty}\!\!\!\!\!dx_1...dx_n\;
f(\Phi_n(\underline{x})) =
\frac{2\,\pi^{n/2}}{(n+1)^{3/2}\,\Gamma(n/2)} \;
\int_0^{+\infty}\!\!\!\!\!dr \,r^{n-1} f(r^2) \; .
\end{equation}

\section{}\label{valeurcn}
\setcounter{equation}{0}

	In this Appendix we derive the explicit form of $I_n(\ell,...,\ell)$ defined
in (\ref{b4}), or equivalently we give the value of the coefficients $C_n$
appearing in Eq.~(\ref{b8bis}). These computations extend the
results of \ref{integrales} (Eqs.~(\ref{b5}) to (\ref{b8})) and are valid for
any $n$. However, they are restricted to the case $\ell_1=...=\ell_n=\ell$.

	 For evaluating the integral (\ref{b4}) it is customary
to make several manipulations~: one performs the $y$
integration in the first term of the r.h.s. In the second term, one can
decrease by one
the number of variables of integration easily, since this term is simply an
elaborate manner of writing $\Theta(y)\,G_0(\vec{r},\vec{r},E)$. After a
scaling on the variables ($y_i=s_i\sqrt{k/(2\ell)}$) one obtains a result
which can be cast in the form~:

\begin{equation}\label{v1}
I_n(\ell,...,\ell) =
\frac{\ep^{ik L +i\pi/2-i\,n\pi/4}}{2\,k\,\pi^{n/2}}\; \tilde{I}_{n-1}
\; ,\end{equation}

\noindent	where

\begin{equation}\label{v2}
\tilde{I}_n = 
\int_0^{+\infty}\!\!\!\!\!dx
\left[\int_{-\infty}^{+\infty}\!\!\!\!\!...\int_{-\infty}^{+\infty}
\!\!\!\!\!dy_1...dy_n\;
\ep^{i\Psi(x,\underline{y})} -
\int_{0}^{+\infty}\!\!\!\!\!
...\int_{0}^{+\infty}\!\!\!\!\!dy_1...dy_n\;
\ep^{i\Psi(x,\underline{y})}
\right]\; ,\end{equation}

\noindent and $\Psi(x,\underline{y})$ is the following quadratic form (we
denote $n$--dimensional vectors $\underline{y} = (y_1,...,y_n)$)

\begin{equation}\label{v3}
\Psi(x,\underline{y})=(x-y_1)^2+(y_1-y_2)^2...+
(y_{n-1}-y_n)^2+(y_n-x)^2
\; .\end{equation}

	$\tilde{I}_n$ in (\ref{v2}) can be expressed simply in term of the function
$\psi_n(x)$ defined by~:
`
\begin{equation}\label{v4}
\psi_n(x)=\int_0^{+\infty}\!\!\!\!\!...\int_0^{+\infty}\!\!\!\!\!
dy_1...dy_n\; \ep^{i\Psi(x,\underline{y})} \; .
\end{equation}

	One first notices that for $x$ large and positive, one can neglect the
boundary effects in the integral (\ref{v4}) defining $\psi_n(x)$ and thus~:

\begin{equation}\label{v5}
\psi_n(x)\underset{x\to +\infty}{\longrightarrow}
\psi_n(+\infty)=
 \int_{-\infty}^{+\infty}\!\!\!\!\!...\int_{-\infty}^{+\infty}\!\!\!\!\!
dy_1...dy_n\; \ep^{i\Psi(x,\underline{y})}
=(\ep^{i\pi/4}\sqrt{\pi})^n\,\frac{1}{\sqrt{n+1}} \; .
\end{equation}

	Hence (\ref{v2}) can be written as~:

\begin{equation}\label{v6}
\tilde{I}_n=\int_0^{+\infty}\!\!\!\!\! dx \, [\psi_n(+\infty)-\psi_n(x)]
= \int_0^{+\infty}\!\!\!\!\! dx \; x\; \psi_n'(x)
\; .\end{equation}

	The function $\psi'_n$ in (\ref{v6}) can be cast in the form

\begin{equation}\label{v7}
\psi_n'(x)=\sum_{m=1}^n \;
\int_{0}^{+\infty}\!\!\!\!\!...\int_{0}^{+\infty}\!\!\!\!\!
dy_1...dy_n \; \delta(y_m) \; \ep^{i\,\Psi(x,\underline{y})}
\; .\end{equation}

	This is done by first changing variables in (\ref{v4}) ($y_j=x+t_j$), then
deriving with respect to $x$ and finally coming back to the original variables
$y_j$. 

	Inserting this expression in (\ref{v6}) and renumbering the variables
in the integral, one obtain the following expression for $\tilde{I}_n$~:

\begin{equation}\label{v9}
\tilde{I}_n = \sum_{m=1}^n \; \langle y_m\rangle \qquad \mbox{where} \qquad
 \langle y_m\rangle
 = \int_{0}^{+\infty}\!\!\!\!\!...\int_{0}^{+\infty}\!\!\!\!\!
dy_1...dy_n \; y_m \,
\, \ep^{\ds i\, \Phi_n(\underline{y})} \; ,\end{equation}

\noindent and the quadratic form $\Phi_n$ is defined in Eq.~(\ref{d2}). The
$n$ integrals $\langle y_m\rangle $ are computed by means of the auxiliary
integral $P_m$ defined as~:

\begin{eqnarray}
P_m & = & -  \int_{0}^{+\infty}\!\!\!\!\!...\int_{0}^{+\infty}\!\!\!\!\!
dy_1...dy_n 
\frac{\partial\,\ep^{\ds i\, \Phi_n(\underline{y})}}{\partial y_m}
=\int_{0}^{+\infty}\!\!\!\!\!...\int_{0}^{+\infty}\!\!\!\!\! 
dy_1...dy_{m-1}dy_{m+1}...dy_n \,
\ep^{\ds i\left.\Phi_n(\underline{y})\right|_{y_m=0}}\label{v10a}\\
& & \nonumber\\
& = & -4\, i \, \langle y_m\rangle +2\, i \,\langle y_{m+1}\rangle 
+2\, i \,\langle y_{m-1}\rangle\label{v10b}
\; .\end{eqnarray}

\noindent with the convention that $\langle y_{-1}\rangle=\langle
y_{n+1}\rangle=0$. The integrals $P_m$ are easily calculated using the results
of \ref{eugene} and noticing that $\left.\Phi_n(\underline{y})\right|_{y_m=0}$
is a sum of two independent quadratic forms $\Phi_{m-1}(y_1,...,y_{m-1})$ and
$\Phi_{n-m}(y_{m+1},...,y_n)$. Therefore the integral of
the r.h.s. of (\ref{v10a}) is the product of two integrals of type (\ref{d1})
(whose explicit form is given in~(\ref{d22}))~:

\begin{equation}\label{v11}
P_m = \tilde{J}_{m-1} \, \tilde{J}_{n-m} = 
\frac{\left(\ep^{i\pi/4}\sqrt{ \pi}\right)^{n-1}}{[m(n-m+1)]^{3/2}} 
\; .\end{equation}

	Then it is a simple matter to solve recursively the system of equations
formed by (\ref{v10b}) and to express the
$\langle y_m\rangle$'s in term of the $P_m$'s. This yields~:

\begin{equation}\label{v12}
\langle y_m\rangle = \frac{m\,\ep^{i\pi/2}}{2(n+1)}
\sum_{q=1}^n (n+1-q)\, P_q
-\frac{\ep^{i\pi/2}}{2}\sum_{q=1}^{m-1}(m-q)\,P_q \; .\end{equation}

\noindent and

\begin{equation}\label{v13}
\tilde{I}_n=\sum_{m=1}^n \; \langle y_m\rangle =\frac{\ep^{i\pi/2}}{4}
\sum_{q=1}^n [q(n-q+1)] \, P_q
\; .\end{equation}

	Using Eq.~(\ref{v11}) one obtains the final formula

\begin{equation}\label{v14}
\tilde{I}_n=\frac{\ep^{i\pi/2}}{4} \, (\ep^{i\pi/4}\sqrt{\pi})^{n-1}\,
\sum_{q=1}^n
\frac{1}{\sqrt{q(n-q+1)}} \; .\end{equation}

	From Eqs.~(\ref{b8bis},\ref{v1},\ref{v14}), the coefficient $C_n$ appearing
in Eq.~(\ref{f7}) reads~:

\begin{equation}\label{v15}
C_n = \frac{1}{8\pi} \sum_{q=1}^{n-1}
\frac{1}{\sqrt{q(n-q)}} \; ,\end{equation}

\noindent which coincides with the results obtained for $n=2,3,4,5$ in
\ref{integrales} by a different method. When $n\to\infty$ the sum over $q$ can
be substituted by an integral and~:

\begin{equation}\label{v16}
\lim_{n\to\infty} C_n = \frac{1}{8\pi} \int_0^n \frac{dx}{\sqrt{x(n-x)}}
=\frac{1}{8}\; .\end{equation}

\end{document}